\documentclass[twocolumn,notitlepage,aps,prb,10pt,superscriptaddress,floatfix,letterpaper,reprint
]{revtex4-2}	

\usepackage{hyperref}
\usepackage[usenames,dvipsnames]{color}
\usepackage{amsmath}
\usepackage[english]{babel}
\usepackage{amssymb}
\usepackage{graphicx}
\usepackage{epstopdf}
\usepackage[normalem]{ulem}
\usepackage{subfigure}
\usepackage{todonotes}
\usepackage{braket}
\usepackage{bbold}
\usepackage{placeins}
\usepackage[para]{threeparttable}
\usepackage{lipsum}
\usepackage{multirow}
\usepackage{bm}
\usepackage{placeins}
\definecolor{linkcolor}{HTML}{399B03}
\definecolor{urlcolor}{HTML}{399B03}
\hypersetup{pdfstartview=FitH, linkcolor=linkcolor,urlcolor=urlcolor,colorlinks=true}
\hypersetup{
    unicode=false,          % non-Latin characters in Acrobat’s  bookmarks
    pdftoolbar=true,        % show Acrobat’s toolbar? 
    pdfmenubar=true,        % show Acrobat’s menu?
    pdffitwindow=false,     % window fit to page when opened
    pdfstartview={FitH},    % fits the width of the page to the window
    pdfauthor={},         % author
    colorlinks=true,        % false: boxed links; true: colored links
    linkcolor=blue,         % color of internal links
    citecolor=red,          % color of links to bibliography
}

\begin{document}

%\preprint{APS/123-QED}

\title{Electron correlations in cubic paramagnetic perovskite Sr(V,Mn)O$_{3}$ - Results from fully self-consistent self-energy embedding calculations.}% Force line breaks with \\

\author{Chia-Nan Yeh}%
\affiliation{%
 Department of Physics, University of Michigan, Ann Arbor, Michigan 48109, USA
}%
\author{Sergei Iskakov}
\affiliation{%
 Department of Physics, University of Michigan, Ann Arbor, Michigan 48109, USA
}%
\author{Dominika Zgid}%
\affiliation{%
 Department of Chemistry, University of Michigan, Ann Arbor, Michigan 48109, USA
}%
\affiliation{%
 Department of Physics, University of Michigan, Ann Arbor, Michigan 48109, USA
}%
\author{Emanuel Gull}%
\affiliation{%
 Department of Physics, University of Michigan, Ann Arbor, Michigan 48109, USA
}%

\date{\today}

\begin{abstract}
In this work, we use the thermodynamically consistent and conserving self-energy embedding theory (SEET) to study the spectra of the prototypical undistorted cubic perovskites SrVO$_3$ and SrMnO$_3$. 
%In the strongly correlated metallic SrVO$_{3}$, local correlations in the V $t_{2g}$ orbitals are overestimated due to the insufficient non-local correlations present in $GW$. 
In the strongly correlated metallic SrVO$_3$ we find that the usual attribution of the satellite peaks at -1.8eV to Hund or Hubbard physics in the $t_{2g}$ orbitals is inconsistent with our calculations. 
In the strongly correlated insulator SrMnO$_3$ we recover insulating behavior due to a feedback effect between the strongly correlated orbitals and the weakly correlated environment.
Our calculation shows a systematic convergence of spectral features as the space of strongly correlated orbitals is enlarged, paving the way to a systematic parameter free study of correlated perovskites.
\end{abstract}

\maketitle

\section{Introduction}
The simulation of strongly correlated solids within a rigorous parameter-free ab-initio theory is one of the big challenges of modern quantum physics. 
The goal is to find approximate methods able to describe the complexities of the electronic structure problem while, at the same time, taking into account electron correlations accurately.

The distortionless cubic oxide perovskites SrVO$_3$ and SrMnO$_3$ form an ideal testbed for these theories~\cite{Pavarini04,SrVO3_Sekiyama04,Lechermann06,SrVO3_Nekrasov05,Nekrasov06,SrVO3_Taranto13,SrVO3_Tomczak12,SrVO3_Sakuma13,Dang14,Chen14,Tomczak14,Bauernfeind17,Bauernfeind18}. Of particular interest is the single-particle excitation spectrum, which can be probed by photemission and inverse photoemission techniques and which, in diagrammatic techniques, is directly related to the imaginary part of the real frequency spectral function.

While the spectrum of both materials is well characterized experimentally, standard electronic structure methods such as the density functional theory (DFT) and GW are not able to reproduce it due to the missing correlations in their partially filled transition metal shells. For instance, in the LDA calculations, the $t_{2g}$ quasiparticle bandwidth in SrVO$_{3}$ is too wide, and SrMnO$_3$ is metallic, rather than insulating. 

Both materials are prototypes for density functional theory plus dynamical mean field theory (DFT+DMFT) simulations~\cite{Pavarini04,SrVO3_Sekiyama04,Lechermann06,SrVO3_Nekrasov05,Nekrasov06,SrVO3_Taranto13,Dang14,Chen14,Bauernfeind17,Bauernfeind18}.
In its simplest variant, this method fits three to five near-Fermi-surface bands and introduces effective Hubbard and Hunds parameters that broaden, split and shift the band structure such that `quasiparticle peaks' and `Hubbard sidebands' can be identified.
In DFT+DMFT, an empirical choice of $U=5$ eV and $J=0.68$ eV generates a characteristic three-peak structure in SrVO$_3$, with a `satellite' peak near $-$1.8 eV and a strong quasiparticle feature, leading to spectra that are remarkably close to experiment and entirely due to Hubbard physics~\cite{Pavarini04,Lechermann06,Bauernfeind17}. 
Sometimes, $U$ and  $J$ parameters determined by constrained LDA (cLDA)~\cite{Gunnarsson89} are used although these parameters could lead to less accurate results~\cite{SrVO3_Sekiyama04,SrVO3_Nekrasov05,Nekrasov06,SrVO3_Taranto13}. 
%Alternatively, $U$ and  $J$ parameters determined by constrained LDA (cLDA)~\cite{Gunnarsson89} could lead to less accurate results~\cite{SrVO3_Sekiyama04,SrVO3_Nekrasov05,Nekrasov06,SrVO3_Taranto13}. 
Similarly, a parameter choice of $U=5$ eV and $J=0.6$ eV along with an extra static double counting shift beyond the fully localized limit (FLL) by $-$2.0 eV turns SrMnO$_3$ into a correlated insulator~\cite{Bauernfeind18}.

More elaborate calculations, such as the ones performed using the $G_{0}W_{0}$+$GW$+EDMFT multitier scheme of Ref.~\cite{Boehnke16,Nilsson17,GW_EDMFT_PRM_Philipp20}, avoid choosing empirical parameter choice by resorting to a parameter-free cRPA~\cite{Aryasetiawan04} calculation together with an extended impurity solution using `screened interactions' in Wannier-downfolded $d$ bands. 
For SrVO$_{3}$, the $G_{0}W_{0}$+$GW$+EDMFT results show that a naive interpretation of the `satellite' peak as Hubbard phenomenon is inconsistent with the calculated spectra and the authors find plasmon physics. However in contrast to the experiment, for SrMnO$_{3}$, in $G_{0}W_{0}$+$GW$+EDMFT the paramagnetic phase (PM) is found to be metallic.

Here, we evaluate the photoemission spectra of both SrVO$_{3}$ and SrMnO$_{3}$ using a parameter-free ab-initio self-energy embedding theory (SEET) without resorting to screened interactions, Wannier function downfolding, or a DFT functional dependent starting point.
We obtain a diagrammatic solution of the entire system in a thermodynamically consistent and conserving fashion, approximating the exact Luttinger-Ward functional $\Phi$ of the solid~\cite{Luttinger60,Baym61,Baym62}. 
Starting from a self-consistent GW solution, we disentangle the individual effects of correlations in physically relevant orbital subgroups by systematically adding correlation diagrams to $\Phi$. We demonstrate that gradually adding non-perturbative terms on the local vanadium (V), manganese (Mn) t$_{2g}$, e$_g$, and oxygen (O) $p$ orbitals leads to a systematic convergence of the results towards photoemission data.

The remainder of this paper will proceed as follows. In Sec.~\ref{sec:methods} we will introduce the method and the computational details. In Sec.~\ref{sec:SrVO3} we will discuss its application to SrVO$_3$ and in Sec.~\ref{sec:SrMnO3} to SrMnO$_3$. In Sec.~\ref{sec:conc} we will present conclusions.

\section{Method}\label{sec:methods}
We study the electronic structure of paramagnetic SrVO$_{3}$ and SrMnO$_{3}$ using SEET~\cite{Kananenka15,Zgid17,Rusakov19} based on the procedures described in Ref.~\cite{Iskakov20}, using \emph{GW}~\cite{Hedin65,Aryasetiawan98,Kutepov09,Boehnke16,Iskakov20} as the weakly correlated `outer' method for all orbitals and Exact Diagonalization (ED)~\cite{Caffarel94,ED_Sergei18} as the `inner' quantum impurity solver for the correlated orbitals. 
We emphasize that, in the present work, we iterate the SEET equations fully to self-consistency~\cite{Rusakov19,Iskakov20} as shown in Fig.~\ref{fig:SEET_workflow} (see Sec.~\ref{subsec:SEET} for details). This ensures that the self-energy in the weakly correlated orbitals is adjusted to reflect the influence of the self-energy coming from the strongly correlated orbitals.
Full self-consistency is essential for a $\Phi$-derivable theory that respects thermodynamic consistency and conservation laws~\cite{Luttinger60,Baym61,Baym62,Iskakov19}, and will be needed to recover the insulating character of the PM phase of SrMnO$_3$.

Our implementation of SEET does not rely on low-energy projections of orbitals and interactions but instead is formulated in terms of the bare electronic structure Hamiltonian.
No effective model parameters, double counting corrections, or other adjustable parameters are employed.
%SEET requires the separation of the system into strongly correlated subspaces, which are treated exactly, and the environment, which is treated here at the level of \emph{GW}. 
Non-local screening is fully included at the level of \emph{GW}.
The method is controlled in the sense that as the correlated subspace is enlarged, all correlations of the electronic structure problem are gradually recovered. 

\subsection{Self-energy embedding theory}\label{subsec:SEET}
We study  the periodic electronic structure Hamiltonian
\begin{align}
H = \sum_{\boldsymbol{ij},\sigma}h^{0}_{\boldsymbol{ij}}c^{\dag}_{\boldsymbol{i}\sigma}c_{\boldsymbol{j}\sigma} + \frac{1}{2}\sum_{\boldsymbol{ijkl},\sigma\sigma'}v_{\boldsymbol{ijkl}}c^{\dag}_{\boldsymbol{i}\sigma}c^{\dag}_{\boldsymbol{k}\sigma'}c_{\boldsymbol{l}\sigma'}c_{\boldsymbol{j}\sigma},
\label{eqn:H}
\end{align} 
where $c^{\dag}_{\boldsymbol{i}\sigma}$ ($c_{\boldsymbol{i}\sigma}$) are creation (annihilation) operators for the single-particle state with spin $\sigma$ and multi-index $\boldsymbol{i}$, which represents orbital $i$ and crystal momentum $\bold{k}_{i}$. 
We use symmetrized atomic orbitals (SAO)~\cite{Lowdin70_SAO} constructed from Gaussian Bloch orbitals as the orbital basis. 
$h^{0}_{\boldsymbol{ij}}$ and $v_{\boldsymbol{ijkl}}$ are the standard single-particle and two-particle operators; for an explicit definition see {\it e.g.} Eq. 4 of Ref.~\cite{Iskakov20}. 
%Translational symmetry is always implicitly applied to both $h^{0}_{\boldsymbol{ij}}$ and $v_{\boldsymbol{ijkl}}$.

As a starting point, we employ the $GW$ approximation as a weakly correlated method to obtain the momentum-resolved Green's functions $(G^{GW})^{\bold{k}}$ and self-energies $(\Sigma^{GW})^{\bold{k}}$. In $GW$, $(\Sigma^{GW})^{\bold{k}}$ is a functional $\mathcal{F}_{GW}$ of $(G^{GW})^{\bold{k}}$, 
\begin{align}
(\Sigma^{GW})^{\bold{k}} = \mathcal{F}_{GW}[(G^{GW})^{\bold{k}}].
\label{eqn:scGW}
\end{align}
%see {\it e.g.} Ref.~\cite{Iskakov20}. 
Along with the Dyson equation 
\begin{align}
(G^{GW}(i\omega_{n}))^{\bold{k}} = \Big [ (i\omega_{n} + \mu) - h^{0,\bold{k}} - (\Sigma^{GW}(i\omega_{n}))^{\bold{k}}\Big ]^{-1}, 
\label{eqn:G_GW}
\end{align}
Eq.~\ref{eqn:scGW} implies a self-consistency between $\Sigma^{GW}$ and $G^{GW}$, defining the so-called self-consistent $GW$ (sc$GW$) approximation.
Corrections to $GW$ occur at second order in the interaction, and therefore the method may become unreliable when applied to strongly interacting systems. 

SEET includes non-perturbative corrections to the $GW$ self-energy diagrams within subsets of potentially strongly correlated orbitals.
Here, we choose as the strongly correlated subsets $M$ disjoint groups of local orbitals close to the Fermi energy $E_{F}$, labeled by $A_1\dots,A_M$.

SEET modifies the self-energy of Eq.~\ref{eqn:scGW}~\cite{Zgid17}
\begin{align}
(\Sigma^{\text{SEET}})^{\bold{k}}_{ij} &=(\Sigma^{\text{weak}})^{\bold{k}}_{ij} \nonumber \\ &+ \sum_{\lambda=1}^M\big[(\Sigma^{\text{non-pert}}_{A_\lambda})_{ij} - (\Sigma^{\text{DC}}_{A_\lambda})_{ij}\big]\delta_{(ij)\in A_\lambda},
\label{eqn:Sigma_SEET_k}
\end{align}
where $(\Sigma^{\text{weak}})^{\bold{k}}$ denotes the perturbatively evaluated weak correlation solution of the entire system, $(\Sigma^{\text{non-pert}}_{A_\lambda})$  the non-perturbatively evaluated contribution within the orbital set $A_\lambda$, $\lambda=1,...,M$, and the double-counting term $(\Sigma^{\text{DC}}_{A_\lambda})$ ensures that no self-energy contribution is counted twice. $\delta_{(ij)\in A_\lambda}$ is $1$ only if both orbital $i$ and orbital $j$ are part of subspace $A_\lambda$, and $0$ otherwise. 
%In this article, we shall denote the difference between $(\Sigma^{\text{SEET}})^{\bold{k}}_{ij\in A_{\lambda}}$ and $(\Sigma^{GW})^{\bold{k}}_{ij\in A_{\lambda}}$ as local self-energy correction to orbital group $A_{\lambda}$. 

The corresponding interacting Green's function is 
\begin{align}
(G^{\text{SEET}}(i\omega_{n}))^{\bold{k}} = \Big [ (i\omega_{n} + \mu) - h^{0,\bold{k}} - (\Sigma^{\text{SEET}}(i\omega_{n}))^{\bold{k}}\Big ]^{-1}.
\label{eqn:G_SEET}
\end{align}

In this work, $(\Sigma^{\text{weak}})^{\bold{k}}$  is evaluated as the $GW$ self-energy with the interacting propagator $(G^{\text{SEET}})^{\bold{k}}$ of the system, $\mathcal{F}_{GW}[(G^{\text{SEET}})^{\bold{k}}]$. 
Similarly, $\Sigma^{\text{DC}}_{A_\lambda}$ is evaluated as the $GW$ self-energy with its vertices restricted to orbital subset $A_\lambda$. 

Choosing the subspace of correlated orbitals local to each unit cell implies that Fourier transforming Eq.~\ref{eqn:Sigma_SEET_k} to real space yields
\begin{align}
(\Sigma^{\text{SEET}})^{\bold{RR'}}_{ij} &= (\Sigma^{\text{weak}})^{\bold{RR'}}_{ij} \nonumber\\
&+ \sum_{\lambda=1}^M\big[(\Sigma^{\text{non-pert}}_{A_\lambda})_{ij} - (\Sigma^{\text{DC}}_{A_\lambda})_{ij}\big]\delta_{(ij)\in A_\lambda}\delta_{\bold{R}\bold{R}'}
\label{eqn:Sigma_SEET_R}
\end{align}
where $\bold{R}$ and $\bold{R}'$ are unit cell indices. 

The Dyson equation then defines the inverse Green's function in orbital space $A_\lambda$,
\begin{align}
(G^{\text{SEET},-1})^{\bold{R}\bold{R}}_{ij\in A_\lambda} &= \Big[(i\omega_{n}+\mu)\mathbf{1} - h^{0,\bold{RR}}_{ij\in A_\lambda} - (\Sigma^{\text{SEET}})^{\bold{RR}}_{ij\in A_\lambda}\Big]. \end{align}
We emphasize that $(G^{\text{SEET},-1})^{\bold{R}\bold{R}}_{ij\in A_\lambda} \neq  [(G^{\text{SEET}})^{\bold{RR}}_{ij\in A_\lambda}]^{-1}$, {\it i.e.}  the inverse Green's function restricted to subset $A_\lambda$ is not equal to the inverse of the Green's function restricted to $A_\lambda$.

The difference between those two quantities,
\begin{align}
&\big[(G^{\text{SEET}})^{\bold{R}\bold{R}}_{ij\in A_\lambda}\big]^{-1} \nonumber\\
&= \Big[(i\omega_{n}+\mu)\mathbf{1} - h^{0,\bold{RR}}_{ij\in A_\lambda} - (\Sigma^{\text{SEET}})^{\bold{R}\bold{R}}_{ij\in A_\lambda}(i\omega_{n}) - \Delta^{A_\lambda}_{ij\in A_\lambda}(i\omega_{n})\Big],
\label{eqn:G_SEET_subA}
\end{align}
defines the hybridization function $\Delta^{A_\lambda}_{ij}$.

\subsection{Impurity Hamiltonian construction}\label{sec:H_imp}
%Note that each self-energy term can be split into the static ($\Sigma_{\infty}$) and dynamic ($\Sigma_{\text{corr}}$) part of self-energy, namely, $\Sigma = \Sigma_{\infty} + \Sigma_{\text{corr}}(i\omega_{n})$. 
By splitting the static and dynamic part of each self-energy contribution ($\Sigma(i\omega_{n}) = \Sigma_{\infty} + \Sigma_{\text{corr}}(i\omega_{n})$), we can reorder terms in Eq.~\ref{eqn:G_SEET} such that 
\begin{align}
\big[(G^{\text{SEET}})^{\bold{R}\bold{R}}_{ij\in A_{\lambda}}\big]^{-1} &= \Big[(i\omega_{n}+\mu)\mathbf{1} - \tilde{h}^{0,\bold{RR}}_{ij\in A_{\lambda}} - \Sigma^{\text{corr},\bold{RR}}_{ij\in A_{\lambda}}(i\omega_{n}) \nonumber\\
&- \Sigma^{\text{non-pert}}_{ij\in A_{\lambda}}(i\omega_{n}) - \Delta^{A_{\lambda}}_{ij\in A_{\lambda}}(i\omega_{n})\Big], 
\label{eqn:G_SEET_ver2}
\end{align}
where $\tilde{h}^{0,\bold{RR}}_{ij\in A_{\lambda}} = h^{0,\bold{RR}}_{ij\in A_{\lambda}} + (\Sigma^{\text{weak}}_{\infty})^{\bold{RR}}_{ij\in A_{\lambda}} - (\Sigma^{\text{DC}}_{A_{\lambda},\infty})_{ij\in A_{\lambda}}$ is the frequency-independent renormalized non-interacting Hamiltonian and $\Sigma^{\text{corr},\bold{RR}}_{ij\in A_{\lambda}} = (\Sigma^{\text{weak}}_{\text{corr}})^{\bold{RR}}_{ij\in A_{\lambda}} - (\Sigma^{\text{DC}}_{A_{\lambda}, \text{corr}})_{ij\in A_{\lambda}}$ is the local dynamic self-energy without double counting contribution. 
In order to solve $\Sigma^{\text{non-pert}}_{A_{\lambda}}$, we define the auxiliary propagator 
\begin{align}
g^{-1}_{A_{\lambda}} = g^{0,-1}_{A_{\lambda}} - \Sigma^{\text{non-pert}}_{ij\in A_{\lambda}},
\label{eqn:g_aux}
\end{align}
where the inverse of the non-interacting auxiliary counterpart $g^{0,-1}_{A_{\lambda}}$ is defined as 
\begin{align}
g^{0,-1}_{A_{\lambda}} = (i\omega_{n} + \mu)\mathbf{1} - \tilde{h}^{0,\bold{RR}}_{ij\in A_{\lambda}} - \Delta^{A_{\lambda}}_{ij\in A_{\lambda}}.
\label{eqn:nonint_g_aux}
\end{align}
The Green's function of Eq.~\ref{eqn:g_aux} can then be obtained by solving an auxiliary quantum impurity problem~\cite{Georges96}
\begin{align}
&H^{A_{\lambda}}_{\text{imp}} = \sum_{ij\in A_{\lambda},\sigma}(\tilde{h}^{0}_{ij\sigma}-\mu\delta_{ij})c^{\dag}_{i\sigma}c_{j\sigma} + \sum_{b,\sigma}\epsilon_{b\sigma} a^{\dag}_{b\sigma}a_{b\sigma} \nonumber\\
&+ \sum_{i\in A_{\lambda},b,\sigma} (V_{ib\sigma}c^{\dag}_{i\sigma}a_{b\sigma} + h.c.) + \frac{1}{2}\sum_{\substack{ijkl\in A_{\lambda}\\ \sigma\sigma'}}v_{ijkl}c^{\dag}_{i\sigma}c^{\dag}_{k\sigma'}c_{l\sigma'}c_{j\sigma}
\label{eqn:H_imp}
\end{align}
within the subset $A_{\lambda}$ where $a^{\dag}_{b\sigma}$ ($a_{b\sigma}$) are creation (annihilation) operators for bath orbital $b$ with spin $\sigma$. The energy levels $\{\epsilon_{b\sigma}\}$ and couplings between bath ($b$) and impurity orbitals ($i$) $V_{ib\sigma}$ can be approximated by discretizing the continuous hybridization function with a finite number $N_{b}$ of bath orbitals
\begin{align}
\Delta^{A_{\lambda}}_{ij\sigma}(i\omega_{n}) = \sum_{b=1}^{N_{b}}\frac{V_{ib\sigma}V_{jb\sigma}}{i\omega_{n}-\epsilon_{b\sigma}}. 
\label{eqn:Delta}
\end{align}

The expressions for $\Sigma$ (as a function of $G$) and $G$ (as a function of $\Sigma$) lead to self-consistent equations that are $\Phi$-derivable and therefore conserving and thermodynamically consistent \cite{Zgid17}. We emphasize that the two-body Coulomb interactions remain the bare interactions of the original electronic Hamiltonian (Eq.~\ref{eqn:H}). Screening effects are addressed by the explicit treatment of $\Sigma^{\text{corr},\bold{RR}}_{ij\in A}$ in Eq.~\ref{eqn:G_SEET_ver2}, and contain all screening contributions on the level of the weak correlation method, as well as non-perturbative correlation effects due to correlations in the subspace $A_{\lambda}$. % the local self-energy. %due to a different definition of hybridization compared to other DMFT-type embedding schemes.
The method is different from DMFT and other embedding methods where the self-consistency condition is different, the local self-energy contains no contributions from the weak coupling terms, and interaction parameters are typically treated as free parameters in the early LDA+DMFT formulations~\cite{RMP_Kotliar06}.

\begin{figure}[tb!]
\includegraphics[width=0.3\textwidth]{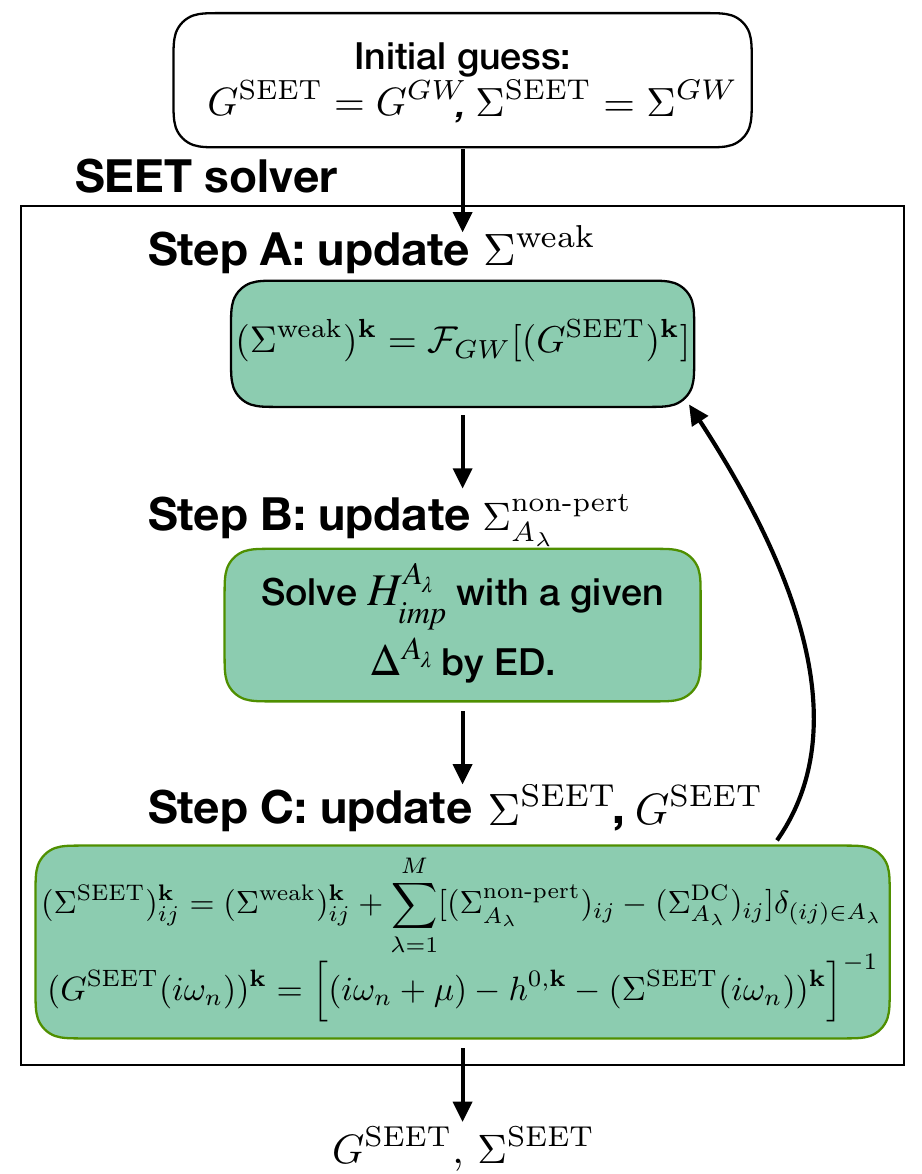}
\caption{Workflow of self-consistent SEET. $H^{A_{\lambda}}_{\text{imp}}$ and $\Delta^{A_{\lambda}}$ are defined in Eq.~\ref{eqn:H_imp} and~\ref{eqn:G_SEET_subA}.}\label{fig:SEET_workflow}
\end{figure}

The SEET equations are solved iteratively until convergence in all quantities is achieved.
The self-consistency loop is illustrated in Fig.~\ref{fig:SEET_workflow}. 
At step A, a single iteration of $GW$, $\mathcal{F}_{GW}[(G^{\text{SEET}})^{\bold{k}}]$, is executed to update $(\Sigma^{\text{weak}})^{\bold{k}}$ for the whole system. 
Along with the definition of hybridization function $\Delta^{A}$ in Eq.~\ref{eqn:G_SEET_subA}, step B corresponds to solving $H^{A_{\lambda}}_{\text{imp}}$ with a quantum impurity solver. 
When multiple subspaces $A_{\lambda}$ are defined, $H^{A_{\lambda}}_{\text{imp}}$ for each subset is solved independently. 
Lastly, $\Sigma^{\text{SEET}}$ and $G^{\text{SEET}}$ are updated according to Eq.~\ref{eqn:Sigma_SEET_k} and ~\ref{eqn:G_SEET} with the new $\Sigma^{\text{weak}}$ and $\Sigma^{\text{non-pert}}_{A_{\lambda}}$. 

Note that the SEET framework can be adapted to use other diagrammatic many-body perturbation approximations as the weakly correlated method, as long as the double counting correction can be rigorously defined. For additional discussion of the method see Ref.~\cite{Zgid17}.

\section{Computational details}\label{sec:comp_details}
We solve the electronic structure Hamiltonian Eq.~\ref{eqn:H} in a Gaussian  \emph{gth-dzvp-molopt-sr} basis~\cite{GTHBasis} with \emph{gth-pbe} pseudopotential~\cite{GTHPseudo} and decompose the four-fermion Coulomb integrals into a combination of auxiliary even-tempered Gaussians for Sr and \emph{def2-svp-ri}~\cite{RI_auxbasis} bases for all other atoms, using up to $6\times6\times6$ k-points in the Brillouin zone. Integrals are obtained with the open source \texttt{PySCF}~\cite{PySCF} package. 

%SEET is based on Green's function formulation where temperature dependency is intrinsically built in. 
All dynamic quantities in SEET are computed on the imaginary time and frequency axis. Efficient representations for both imaginary time and frequency grids are essential for realistic material calculations. 
In this work, we use the compact intermediate representation (IR)~\cite{Shinaoka17} with sparse frequency sampling~\cite{Li20} for all dynamical quantities such as Green's function and self-energy. 
In IR, the grid size is governed by a dimensionless parameter $\Lambda$ that should be at least larger than $\beta\omega_{\text{max}}$ where $\beta$ is the inverse temperature and $\omega_{\text{max}}$ is the energy bandwidth of the system. 
Lower temperature and larger energy bandwidth would require larger grid size and thereby increase computational linear with $\beta$. 
Simulations are performed at temperature $T\sim 1579K$ ($\beta = 200$ Ha$^{-1}$) for SrVO$_{3}$ and $1053$ K ($\beta = 300$ Ha$^{-1}$) for SrMnO$_{3}$.

We choose the strongly correlated subspace to be the local transition metal $3d$ orbitals (split into $t_{2g}$ and $e_{g}$) as well as the oxygen $p$ orbitals. 
Note that standard DMFT impurity constructions typically include $t_{2g}$~\cite{Pavarini04,SrVO3_Sekiyama04,Lechermann06,SrVO3_Nekrasov05,Nekrasov06,SrVO3_Taranto13,SrVO3_Tomczak12,SrVO3_Sakuma13,Tomczak14,Boehnke16,Bauernfeind17,Bauernfeind18,GW_EDMFT_PRM_Philipp20} or $t_{2g}+e_{g}$~\cite{Bauernfeind17,Bauernfeind18,GW_EDMFT_PRM_Philipp20} orbitals, though methods to treat entire unit cells have recently been pioneered~\cite{zhu2020ab}. 
In order to gradually improve our solution, we examine the effect of also treating the correlations on the oxygen $p$ orbitals exactly. 
In order to make impurity size feasible for ED calculations, we further split the O $2p$ shell into two $2p_{\pi}$ and one $2p_{\sigma}$ orbitals, defined with respect to the transition metal ion.

\section{Strontium Vanadate}\label{sec:SrVO3}
\begin{figure*}[tbh]
\includegraphics[width=0.47\textwidth]{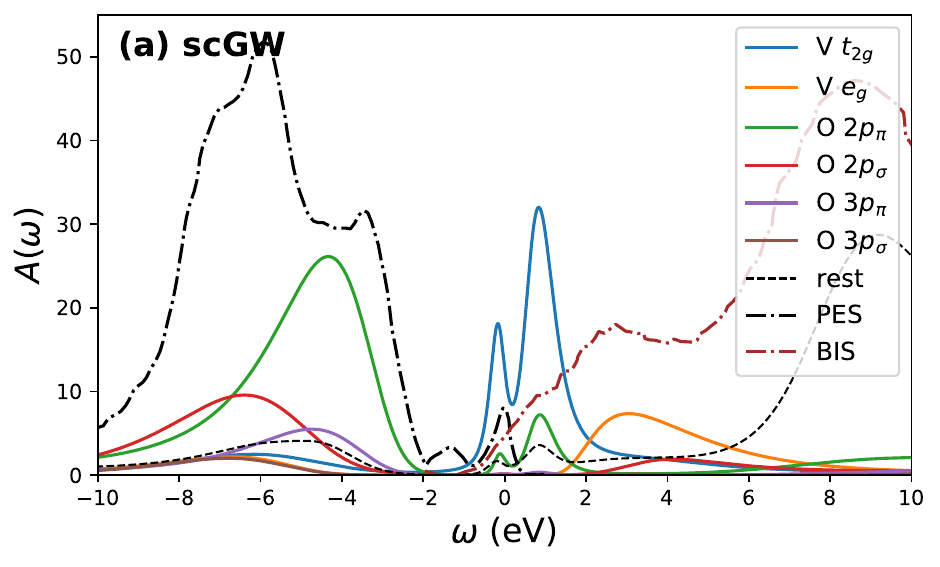}
\includegraphics[width=0.47\textwidth]{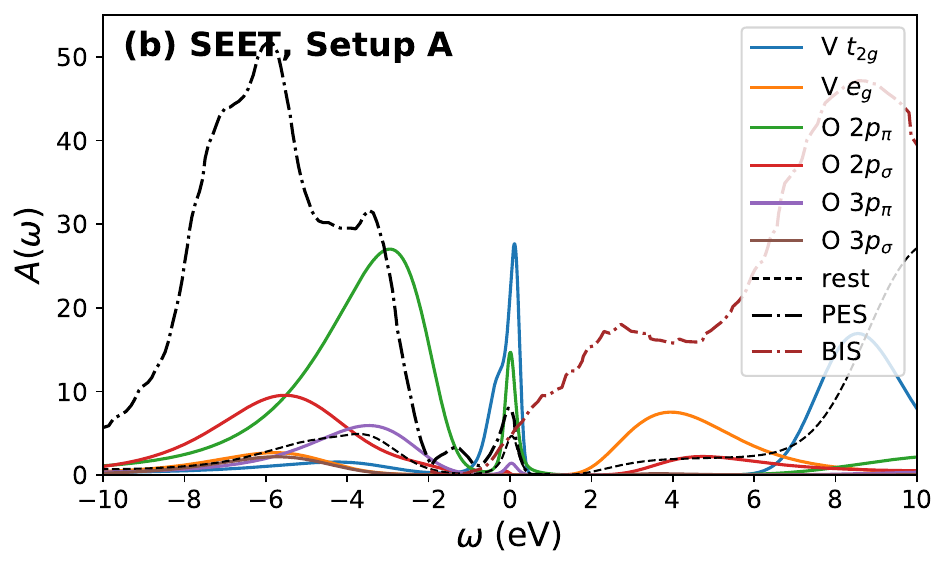} \\
\includegraphics[width=0.47\textwidth]{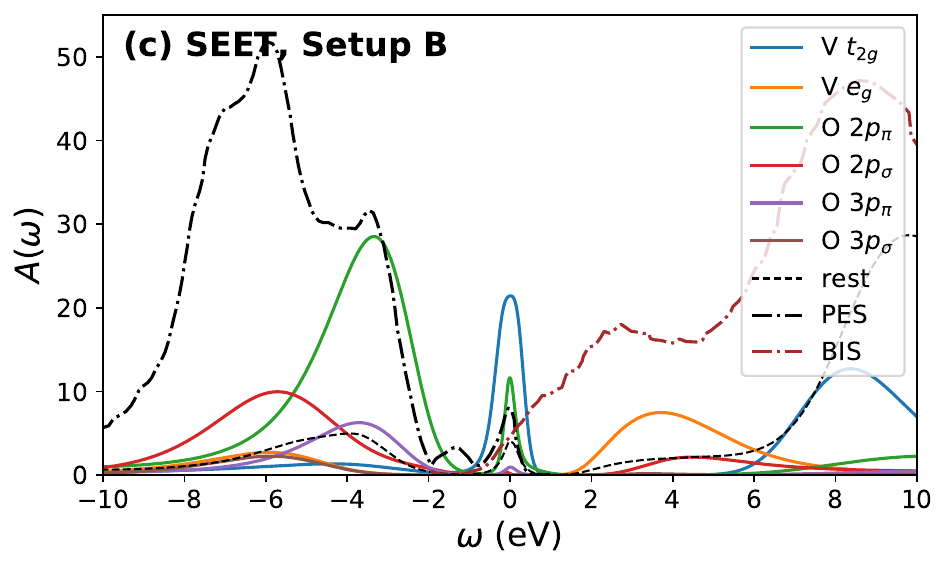} 
\includegraphics[width=0.47\textwidth]{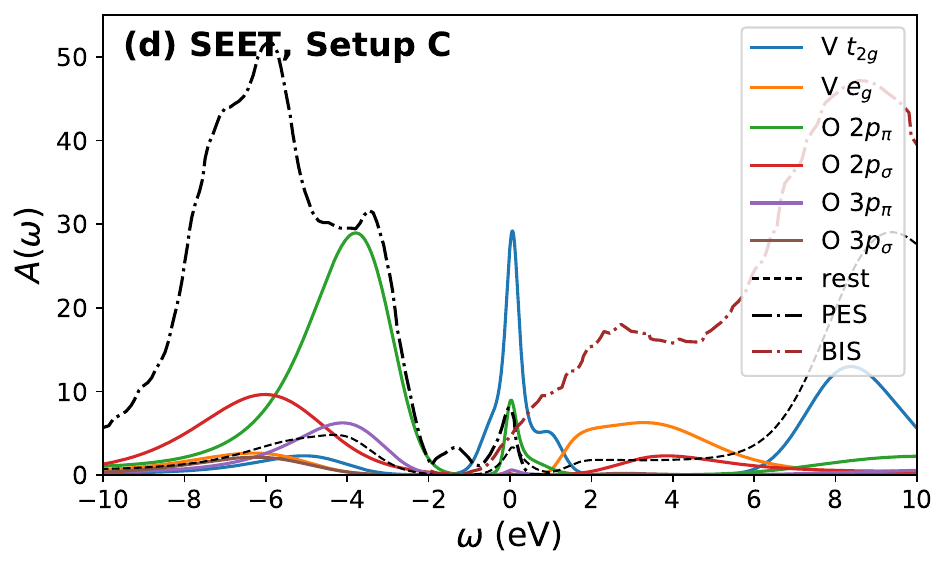} 
\caption{Local orbital-resolved DOS of SrVO$_{3}$ from sc\emph{GW} and SEET. Impurity choices are shown in Table \ref{tab:impurities_SrVO3}. The PES and BIS data are obtained from Ref.~\cite{SrVO3_PES_Yoshimatsu10,SrVO3_Morikawa95}. The experimental data are rescaled to match results from analytical continuation.}\label{fig:PDOS_SrVO3}
\end{figure*}

\begin{table}[bth]
\begin{ruledtabular}
\begin{tabular}{c|c|p{6cm}}
Name & Imp & Description \\
\hline
A & 1 & V $t_{2g}$\\
B & 3 & V $t_{2g}$; O $p_{\pi}$; O $p_{\sigma}$\\
C & 4 & V $t_{2g}$; V $e_{g}$; O $p_{\pi}$; O $p_{\sigma}$ 
\\
\end{tabular}
\end{ruledtabular}
\caption{Choice of the impurities for SrVO$_{3}$. Imp denotes the number of distinct disjoint impurity problems. \label{tab:impurities_SrVO3}}
\end{table}

SrVO$_{3}$ is a simple correlated metal with an undistorted cubic perovskite structure. Nominally, it has a single electron in the vanadium $d$-shell. Photoemission spectroscopy (PES) and bremsstrahlung isochromat spectroscopy (BIS) show a pronounced renormalized quasiparticle peak at the  Fermi level ($E_{F}$), a weak satellite peak at $\sim$ $-$1.8 eV, and a strong satellite peak at $\sim$ 3 eV~\cite{SrVO3_Morikawa95,SrVO3_Sekiyama04,SrVO3_PES_Yoshimatsu10}.
V $2p-3d$ resonance photoemission further attributes V $3d$ character both to the quasiparticle peak at $E_{F}$ and to the lower satellite feature at around $-$1.8 eV~\cite{SrVO3_Sekiyama04,SrVO3_PES_Yoshimatsu10}.
Features between $-$3 eV and $-$10 eV are mainly attributed to the O $p$ states~\cite{Takizawa09,SrVO3_Morikawa95,SrVO3_Sekiyama04,SrVO3_PES_Yoshimatsu10}.

In Fig.~\ref{fig:PDOS_SrVO3}, for SrVO$_{3}$, we show a calculated local single particle spectral function $A(\omega)$ as a function of frequency $\omega$.

This spectral function was analytically continued \cite{Jarrell96,Levy17} from the imaginary to the real axis and it was orbitally  resolved to display V $t_{2g}$(blue), V $e_{g}$(orange) and O $2p$ and $3p$ orbitals. Contributions from all other orbitals are included as ``rest".
The experimental results for PES (Ref.~\cite{SrVO3_PES_Yoshimatsu10}) and for BIS (Ref.~\cite{SrVO3_Morikawa95}) are also plotted for easy comparison.

sc\emph{GW} results show a clear $p-d$ splitting. Quasiparticle peaks around $E_{F}$ are dominated by V $t_{2g}$ while the density of state (DOS) between $-$2.5 and $-$10 eV is dominated by O $p$ orbitals with the first and second peak mainly corresponding to O $p_{\pi}$ and O $p_{\sigma}$. 
The conduction band peak around 3 eV is interpreted as V $ e_{g}$ orbitals, rather than an upper Hubbard band from $t_{2g}$ orbitals. 
Except for the missing lower satellite peak around $\sim$ $-$1.8 eV, sc\emph{GW} results qualitatively agree with photoemission spectroscopy. 
In contrast to the experimental quasiparticle peak at $E_{F}$, sc\emph{GW} has a much wider feature (note that the k-space discretization artificially introduces a double peak structure), consistent with a spectrum calculated with non-local \emph{GW} self-energy in Ref.~\cite{Miyake13}. 
%In addition, the small bump at $-$1.8 eV is absent and O $p$ peaks are shifted upward by from $-$3 eV to $-$4 eV~\cite{Takizawa09}.
In addition, the O $p$ peaks are shifted upward from $-$3 eV to $-$4 eV~\cite{Takizawa09}.

Next, we discuss the self-consistent SEET embedding construction with local V $t_{2g}$ states included non-perturbatively in an impurity (SEET, setup A). 
In contrast to sc$\emph{GW}$ in SEET with setup A, the width of the $t_{2g}$ near-Fermi-energy peak is reduced to 1.5 eV which is much closer to experimental data. 
The oxygen $p$ states are shifted upward in energy and there is a large gap between unoccupied $t_{2g}$ and $e_{g}$ states. However, the oxygen shift is inconsistent with PES and the $t_{2g}-e_{g}$ splitting is inconsistent with BIS.

SEET with setup B adds strong correlations to oxygen $2p$ orbitals in addition to the V $t_{2g}$. 
Since the three oxygen atoms in the unit cell are identical, we only consider the $2p$ orbitals from one oxygen atom. Non-perturbative corrections to the other two oxygen atoms are included by symmetry. 
We observe a shift of the oxygen $p$ states back to lower energies but little change for other orbitals.

Adding correlation from V $e_{g}$ orbitals in SEET with setup C moves the $e_{g}$ conduction bands towards lower energy leading to a spectrum consistent with BIS data. In this setup the oxygen $p$ states are shifted to the location found in PES, recovering the pronounced $p-d$ splitting. However, the so called lowered satellite photoemission peak at $-$1 to $-$2 eV is not recovered. 
The lack of this peak in SEET may be due to several reasons. First, the Gaussian basis set employed may not contain a good description of orbitals necessary to evaluate this peak. However, we find this possibility unlikely since SEET calculations performed in a larger basis set still cannot recover this lower satellite peak. (For details see appendix).
Second, correlations beyond GW that have not been considered here such as strong correlations in higher orbitals (e.g. $4d$) or cross-correlations between disjoint impurities (such as $p-d$ correlations) may be responsible for arising of this peak. These correlations beyond $GW$ could be added to SEET if a larger impurity combing $p-d$ orbitals or containing $4d$ orbitals would be considered. Lastly, this lower satellite peak may be caused by an insufficiency of the $GW$ description for multiple orbitals that cannot be simply contained in SEET as multiple larger impurity problems. 
This lower satellite peak is observed in multitier GW+EDMFT calculations performed in Ref.~\cite{GW_EDMFT_PRM_Philipp20} that attributes it to non-local Coulomb interaction. 
%This lower satellite peak is observed in mulititier GW+EDMFT calculations performed in Ref.~\cite{GW_EDMFT_PRM_Philipp20} that attributes it to non-local correlations beyond GW.

\begin{figure}[tb!]
\includegraphics[width=0.47\textwidth]{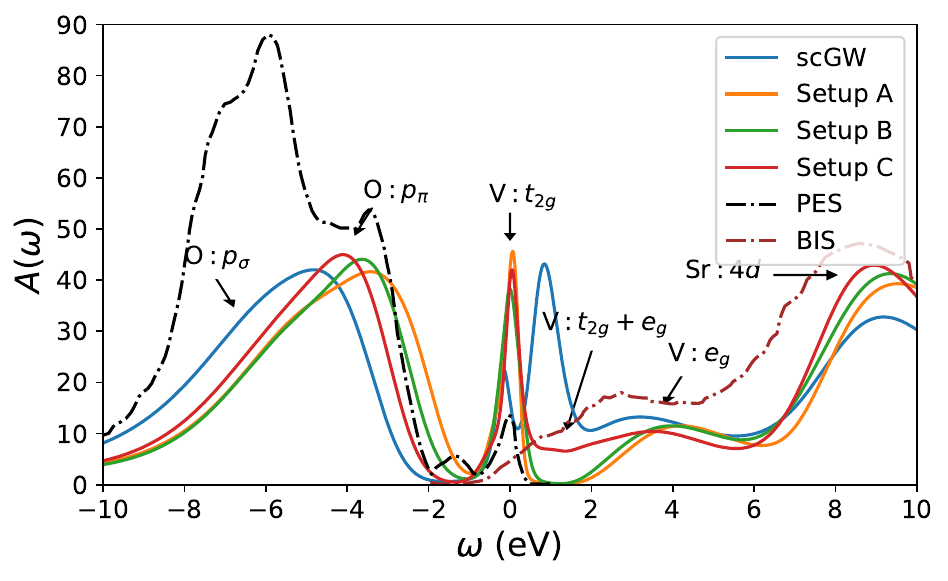}
\caption{Total local density of states of SrVO$_{3}$ from sc\emph{GW} and SEET with different impurity choices. Arrows indicate the orbital contributions according to SEET with impurity choice C. The PES and BIS data is obtained from Ref.~\cite{SrVO3_PES_Yoshimatsu10,SrVO3_Morikawa95}}\label{fig:orbsum_DOS_SrVO3}
\end{figure}

The total local DOS for SrVO$_{3}$ is shown in Fig.~\ref{fig:orbsum_DOS_SrVO3}. 
As we systematically add correlations, SEET reaches quantitatively better agreement with the experimental data. 
The quasiparticle peak around $E_{F}$ is dominated by V $t_{2g}$ while the DOS between $-$2.5 and $-$10 eV is dominated by O $p$ orbitals, with the first and second peak mainly corresponding to O $p_{\pi}$ and O $p_{\sigma}$. 
In the conduction bands, the shoulder at $\approx$ 1 eV and the peak at $\sim$ 2.5$-$3 eV are assigned to V $t_{2g}+e_{g}$ and V $e_{g}$ respectively while the peak at $\approx$ 8 eV is dominated by Sr $4d$ orbitals.
The fact that sc$GW$ predicts a better $p-d$ splitting compared to SEET with setup A implies that there may be error cancellation to the missing local $t_{2g}$ correlation beyond $GW$. 
In general, it is difficult to precisely determine the source of this cancelation since it could be non-local correlations beyond $GW$ from any orbital (e.g. O $p$) or local correlations beyond $GW$ from orbitals other than V $t_{2g}$. 
However, as shown in Fig.~\ref{fig:PDOS_SrVO3} and~\ref{fig:orbsum_DOS_SrVO3}, the position of the O $p$ bands gradually reaches better agreement with experiment from SEET with setup B and C as local self-energy correction of O $2p$ and V $e_{g}$ are added. 
We believe the error cancellation between strong local correlation from V $3d$ and O $2p$ are the reason that sc$GW$ outperforms SEET with setup A. 

\begin{figure}[tbh]
\includegraphics[width=0.45\textwidth]{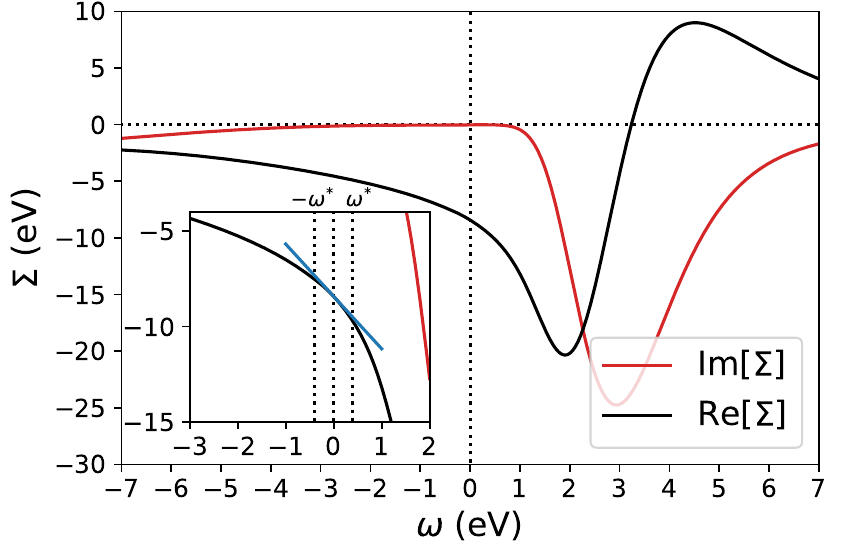}
\caption{V $t_{2g}$ local self-energy from SEET with impurity setup C, analytically continued to the real frequency axis. The estimated quasiparticle regime is around $[-\omega^{*},\omega^{*}]$ where $\omega^{*}=0.4$ eV. }\label{fig:Selfenergy_real_axis}
\end{figure}

In Fig.~\ref{fig:Selfenergy_real_axis}, we show the SEET (setup C) V $t_{2g}$ local self-energy, analytically continued from the imaginary to the real frequency axis. 
The strong pole at around 3.5 eV is consistent with the absence of a V $t_{2g}$ upper Hubbard band at around 3.0 eV, and shifts additional V $t_{2g}$ features  near 8 eV.
The quasiparticle renormalization factor $Z$ can be computed through $Z^{-1} = [1 - \partial \text{Re}\Sigma(\omega)/\partial \omega\Big|_{\omega = 0}]$ for sc$GW$ ($Z=0.7$) and SEET with impurity setup C ($Z=0.27$). In the presence of local self-energy corrections to the V $t_{2g}$ orbitals, the V $t_{2g}$ quasiparticle peak has been strongly renormalized. The reduction of $Z$ from 0.7 to 0.27 is smaller than the experimental value of $Z = 0.5\sim0.6$~\cite{SrVO3_Sekiyama04,SrVO3_Yoshida05} and data from other theoretical calculations~\cite{SrVO3_Sekiyama04,SrVO3_Nekrasov05,Nekrasov06,SrVO3_Taranto13}. 
%The system is predicted to be too correlated by SEET in the presence of non-local screening from $GW$. 
Note that our $Z$ is similar to that of LDA+DMFT (see the inset of Fig. 5 in Ref.~\cite{Nekrasov06}) if the self-energy derivative is taken over the same interval.

Typical failures of LDA for SrVO$_{3}$ include 
(i) a $t_{2g}$ quasiparticle bandwidth that is too large,
(ii) missing lower/upper satellite peaks, and 
(iii) an incorrect position of orbitals outside the V $3d$ shell~\cite{Pavarini04,SrVO3_Sekiyama04,Lechermann06,SrVO3_Nekrasov05,Nekrasov06,SrVO3_Taranto13}.

In the LDA+DMFT scheme, the Coulomb interaction $U$ and the Hund's coupling $J$ are usually chosen \emph{ad hoc} such that a lower/upper Hubbard band emerges~\cite{Pavarini04,Lechermann06,Bauernfeind17}. Sometimes $J$ determined from Hartree-Fock calculations on model Hamiltonians~\cite{Mizokawa96} is chosen. 
A double counting shift of the chemical potential is chosen such that the number of electrons in the DMFT is fixed ($3d^{1}$ configuration in this case), assuming that the strongly correlated subspace is well separated from all other orbitals.
With these parameters, LDA+DMFT produces a quasiparticle band narrowing and a broad satellite at around $-$1.8 eV~\cite{Pavarini04,Lechermann06}.
Less \emph{ad hoc} parameter choices have also been considered but result in less accurate results. 
For example, $U$ and $J$ determined by constrained LDA (cLDA)~\cite{Gunnarsson89} place the lower Hubbard band too low in energy (between $-$2.0 to $-$2.3 eV)~\cite{SrVO3_Sekiyama04,SrVO3_Nekrasov05,Nekrasov06,SrVO3_Taranto13} 
and the FLL double counting~\cite{FLL_94} results in an incorrect $p-d$ splitting in the occupied bands~\cite{Dang14,Chen14}.
In LDA+DMFT, the lower satellite is interpreted as a Hubbard band due to strong local correlations on the vanadium atoms. 

In addition to LDA+DMFT, variants of $G_{0}W_{0}$+DMFT have been applied to SrVO$_{3}$~\cite{SrVO3_Tomczak12,SrVO3_Taranto13,SrVO3_Sakuma13,Tomczak14}. In this scheme, $U(\omega)$ and $J(\omega)$ are determined through constrained RPA (cRPA)~\cite{Aryasetiawan04} based on $G_{0}W_{0}$ and the double counting correction has a rigorous definition if self-consistency is achieved, unlike in LDA+DMFT. In general, $G_{0}W_{0}$+DMFT predicts qualitatively similar results to LDA+DMFT except for the wider $t_{2g}$ quasiparticle bandwidth~\cite{SrVO3_Sakuma13} and better agreement of the lower satellite feature with photoemission data~\cite{SrVO3_Taranto13}. 
As in LDA+DMFT, the lower satellite peak is caused by Hubbard physics.

Recent studies from multitier \emph{GW}+EDMFT found screening beyond \emph{GW} in the V $t_{2g}$ orbitals. 
These effects are resulted from (i) retarded on-site interactions in EDMFT and (ii) local vertex corrections to the polarization~\cite{Boehnke16,Nilsson17,GW_EDMFT_PRM_Philipp20}. 
Note that during the self-consistency loop of multitier $GW$+EDMFT, local vertex corrections to the polarization are involved in the evaluation of the non-local screened interaction. 
The resulting renormalized screened interaction therefore contains more non-local correlations than the one from $GW$. 
%Recent studies from multitier GW+ EDMFT found significant non-local screening in the V t2g-orbitals caused by effects outside of GW~\cite{Boehnke16,Nilsson17,GW_EDMFT_PRM_Philipp20}. 
These results attribute the lower satellite feature to a non-local plasmon, rather than Hubbard physics.
This interpretation is also consistent with the findings from the cumulant expansion based on quasiparticle $GW$~\cite{SrVO3_Gatti13}. 

Although cross-correlations or contributions from higher states, as well as the non-local plasmonic physics examined in multitier $GW$+EDMFT, are outside the physics investigated here, our calculations offer additional insight into the origin of the lower satellite peak at around $-$1.8 eV. 
While Ref.~\cite{Boehnke16,Nilsson17,GW_EDMFT_PRM_Philipp20} include only V $t_{2g}$ and $e_{g}$ orbitals in the correlated subspace, we find that O $2p$ orbitals do not contribute to  this peak. 
Instead, the strong local correlations from O $2p$ improve the the $p-d$ splitting in the presence of non-perturbative V $t_{2g}$ local correlations. 

%Our calculations offer insight into this controversy. 
%First, our results are in agreement with multitier $GW$+EDMFT~\cite{Boehnke16,Nilsson17,GW_EDMFT_PRM_Philipp20} suggesting that the satellite peak at $-$1.8 eV is not caused by local Hubbard physics from the vanadium $3d$ $t_{2g}$ or $e_{g}$ orbitals. 
%While Ref.~\cite{Boehnke16,Nilsson17,GW_EDMFT_PRM_Philipp20} includes only V $t_{2g}$ and $e_{g}$ orbitals in the correlated subspace, we also examined that an O $p$ contributions do not contribute to the rise of this peak. However, cross-correlations or contributions from higher states, as well as the non-local plasmonic physics examined in multitier GW+EDMFT, are outside the physics investigated here, and a clear attribution is not possible within SEET for the correlated subspaces chosen in this paper.

A second observation concerns the contribution of the experimentally observed incoherent feature around 3.0 eV~\cite{SrVO3_Morikawa95} that is traditionally attributed to the upper Hubbard band of the $t_{2g}$ orbitals and $e_{g}$ orbitals.
In our calculation, this peak has only $e_{g}$ character, consistent with some of the literature~\cite{SrVO3_Tomczak12,SrVO3_Gatti13,Tomczak14}.  

%In our calculation, this peak has clear $e_{g}$ character, again inconsistent with a Hubbard interpretation of near-Fermi-surface physics~\cite{Pavarini04,SrVO3_Sekiyama04,SrVO3_Nekrasov05,Nekrasov06,Lechermann06,SrVO3_Taranto13,SrVO3_Sakuma13} but consistent with other approximations~\cite{SrVO3_Tomczak12,SrVO3_Gatti13,Tomczak14}.

\section{Strontium Manganate}\label{sec:SrMnO3}
\begin{figure*}[tbh]
\includegraphics[width=0.47\textwidth]{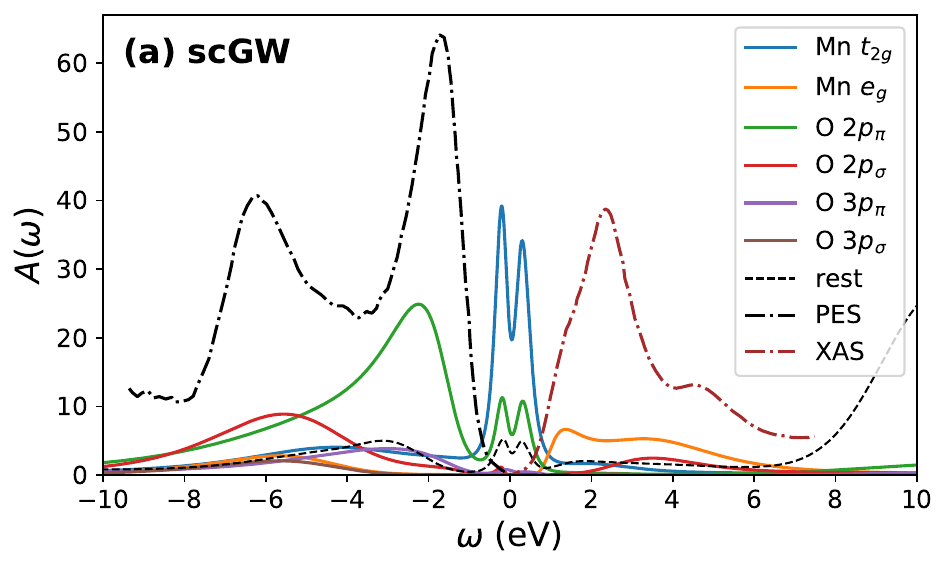}
\includegraphics[width=0.47\textwidth]{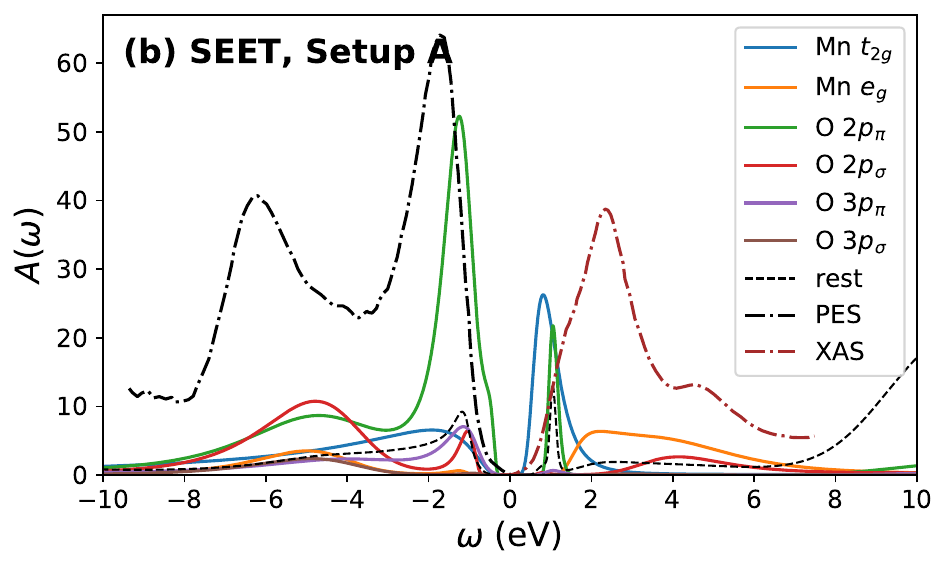}\\
\includegraphics[width=0.47\textwidth]{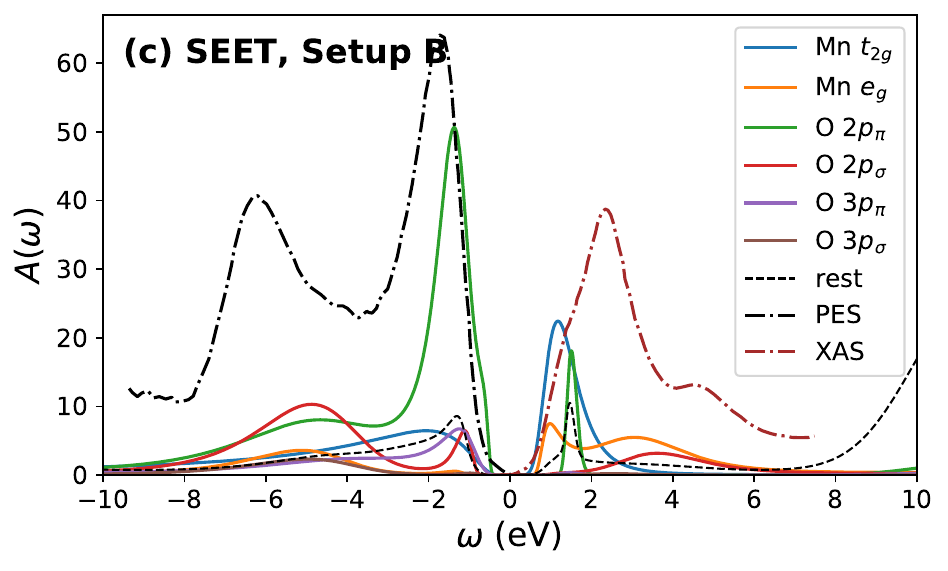}
\includegraphics[width=0.47\textwidth]{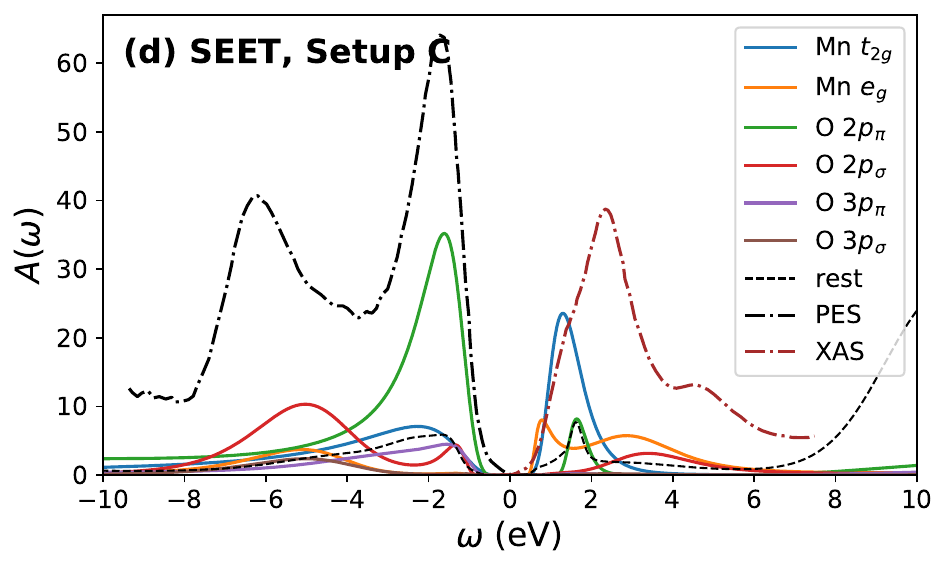}
\caption{Local orbital-resolved DOS of SrMnO$_{3}$ from sc\emph{GW} and SEET. The impurity choices correspond to (A), (B), and (C) in Table \ref{tab:impurities_SrMnO3}. Dotted lines are PES and soft x-ray absorption spectroscopy (XAS) from Ref.~\cite{SrMnFeO3_expt_Kim10}.}\label{fig:PDOS_SrMnO3}
\end{figure*}

\begin{table}[tbh]
\begin{ruledtabular}
\begin{tabular}{c|c|p{6cm}}
Name & Imp & Description \\	
\hline
A & 1 & Mn $t_{2g}$\\
B & 2 & Mn $t_{2g}$; Mn $e_g$ \\
C & 4 & Mn $t_{2g}$; Mn $e_g$; O $p_{\pi}$; O $p_{\sigma}$ 
\\
\end{tabular}
\end{ruledtabular}
\caption{Choice of the impurities for SrMnO$_{3}$. Imp denotes the number of distinct disjoint impurity problems. \label{tab:impurities_SrMnO3}
}
\end{table}

SrMnO$_{3}$ is a cubic insulating perovskite with nominal filling of three electrons in the Mn $3d$ shell. 
Its magnetism is experimentally observed as $G$-type antiferromagnetic (AFM) ordering at low temperature and PM ordering at high temperature (with N$\acute{\text{e}}$el temperature $T_{N}$ $\sim$ 233$-$260 K)~\cite{Negas70,Takeda74}. 
The paramagnetic state has been studied in photoemission~\cite{Abbate92,Saitoh95,Kang08,SrMnFeO3_expt_Kim10} and gap values ranging from 1.0$-$2.3 eV have been found.
To our knowledge, a detailed experimental analysis of the orbital character of the near-Fermi-surface states has not been performed.
Theoretical calculations so far are limited to the AFM state~\cite{Rune06,Dang14,GW_EDMFT_PRM_Philipp20}.

Fig.~\ref{fig:PDOS_SrMnO3} shows the total local orbital-resolved DOS from sc\emph{GW} and SEET with different impurity choices in the PM phase at 1053 K (see Table.~\ref{tab:impurities_SrMnO3}).
In \emph{GW}, SrMnO$_{3}$ is incorrectly predicted to be metallic. 
The corresponding spectral functions are similar to SrVO$_{3}$, where bands around $E_{F}$ are dominated by transition metal $t_{2g}$ states. However, hybridizations between Mn $3d$ and O $2p$ are much stronger, pushing the O $2p$ bands closer to $E_{F}$. 
This strong hybridization puts  the validity of impurity models with only Mn $3d$ orbitals in doubt.
The qualitative failure of sc\emph{GW} implies the need of self-energy diagrams beyond $GW$ approximation.

We first include local self-energy corrections within only the Mn $t_{2g}$ orbitals as shown in Setup A of Fig.~\ref{fig:PDOS_SrMnO3}. 
The non-perturbative treatment of Mn $t_{2g}$ greatly suppress the DOS at $E_{F}$ and opens the gap for SrMnO$_{3}$ which is formed by Mn $t_{2g}$ +  O $p_{\pi}$ hybridized orbitals both above and below the gap. However, the gap is substantially smaller than the one found in photoemission data~\cite{SrMnFeO3_expt_Kim10}. 

Next, we include Mn $e_g$ orbitals in setup B. The non-perturbative treatment of the Mn $e_g$ orbitals results in the conduction band peak of the $e_{g}$ orbitals being pushed below the $t_{2g}$ peak and the gap edge aligning.
Most of the higher valence band states have O $2p_{\pi}$ character, while most of the lower conduction band states have Mn $t_{2g}$ and Mn $e_{g}$ character.

Finally, we include O $2p$ orbitals in impurity setup C. 
Similar to SrVO$_{3}$, only the $2p$ orbitals from one oxygen atom are considered. 
Local self-energy correction beyond $GW$ from O $2p$ further push both Mn $t_{2g}$ and O $p_{\pi}$ away from $E_{F}$. 
Besides that, there is little change to the remaining orbitals.

\begin{figure}[tb]
\includegraphics[width=0.47\textwidth]{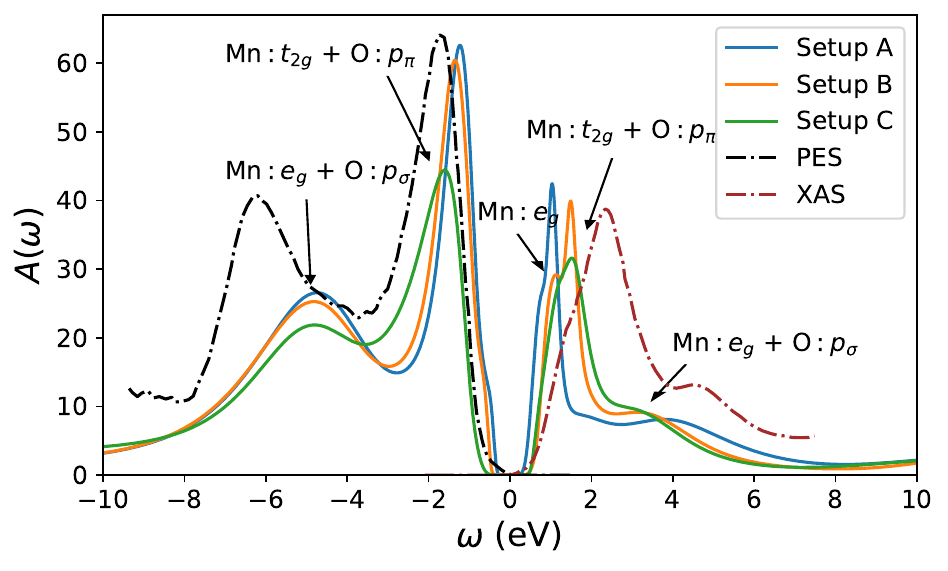}
\caption{Sum of Mn $3d$ + O $2p$ local DOS from SEET with different impurity choices. Dotted lines are photoemission data from Ref.~\cite{SrMnFeO3_expt_Kim10}.}\label{fig:Mn3d_O2p_vs_Expt}
\end{figure}

SEET successfully opens a gap for all impurity choices shown in Fig.~\ref{fig:Mn3d_O2p_vs_Expt}. 
Note that this gap is opened due to the full self-consistency between high and low levels loops in SEET. It only opens when GW orbitals are updated at every iteration to include strong correlation effects that arise from the solution of the impurity problems. We illustrate the opening of the gap versus iteration number as well as the exact convergence for electronic energies in the appending.
Systematic enlarging of the number of impurities leads to a convergence to the experimental data. The first and second valence peaks at $-$2.0 and $-$6.5 eV correspond to Mn $t_{2g}$ + O $p_{\pi}$ and Mn $e_{g}$ + O $p_{\sigma}$. 
In the conduction band, the shoulder at 1 eV corresponds to Mn $e_{g}$ bands, and the two peaks at $-$2.5 and $-$4.8 eV corresponds to Mn $t_{2g}$ + O $p_{\pi}$ and Mn $e_{g}$ + O $p_{\sigma}$. The gap is formed by Mn $t_{2g}$ + O $p_{\pi}$ valence band and Mn $e_{g}$ conduction bands.

The agreement with PES and XAS is not perfect. 
While such a difference may arise due to the finite size effects present, we eliminated this possibility by caring out calculations for different numbers of k-points. For details see appendix. 
Since finite size effects are negligible, the differences are likely due to either 
(i) inter-orbital correlations between different independent impurities, 
and/or (ii) non-local correlations outside of \emph{GW}, (iii) employed Gaussian basis set.

\begin{figure*}[tbh]
\includegraphics[width=0.43\textwidth]{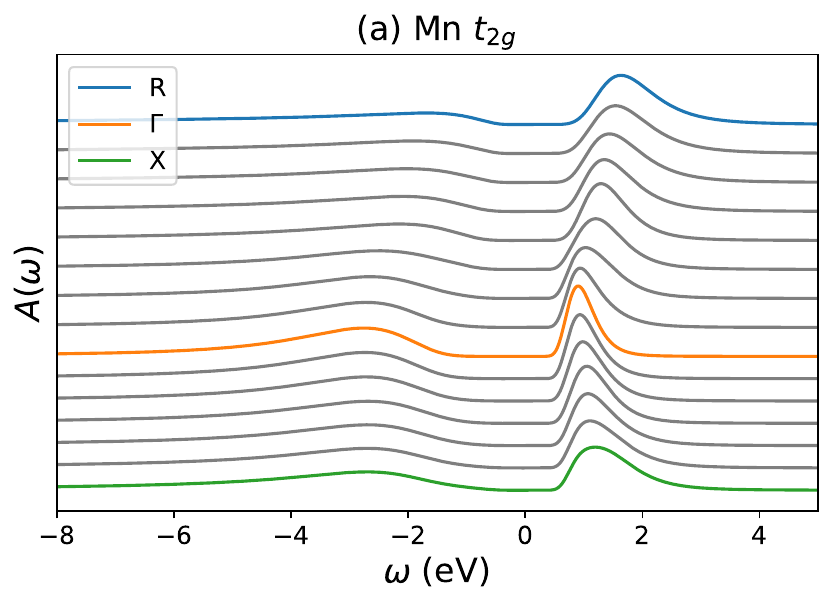}
\includegraphics[width=0.43\textwidth]{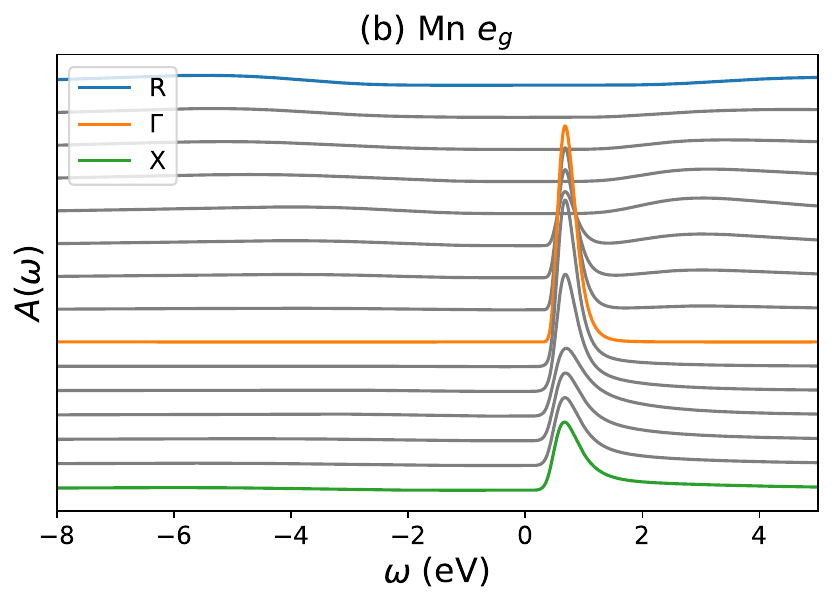}\\
\includegraphics[width=0.43\textwidth]{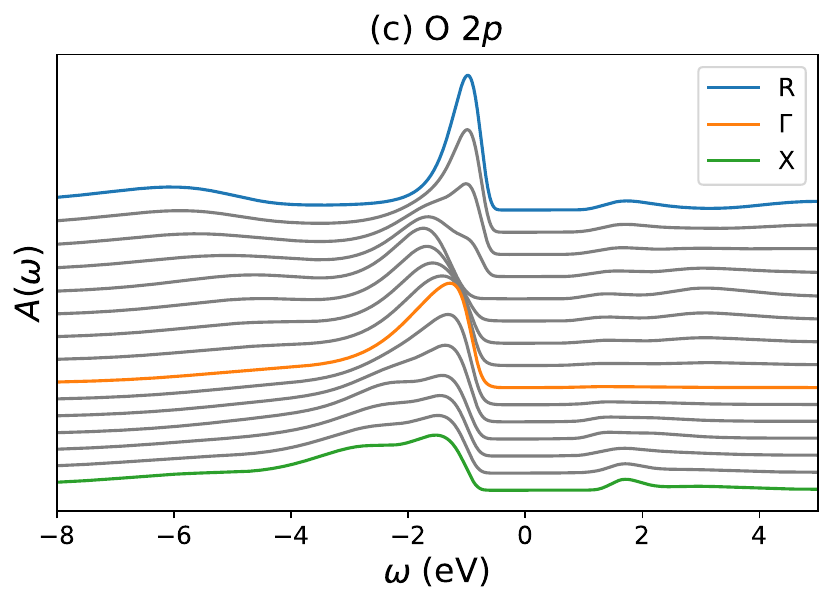}
\includegraphics[width=0.43\textwidth]{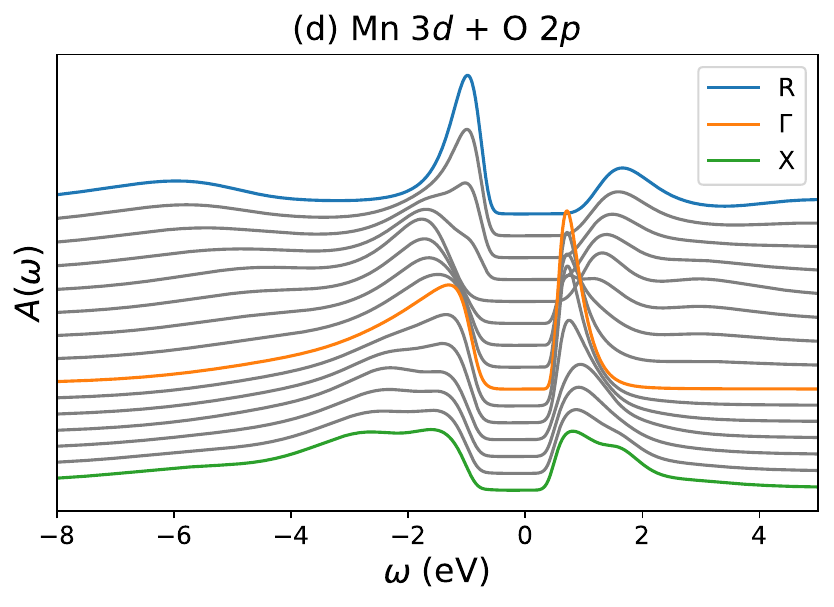}
\caption{Momentum and orbital-resolved partial density of states $A(\omega)$ as a function of frequency $\omega$ of SrMnO$_{3}$ from SEET with impurity setup C. Shown are Mn $t_2g$, Mn $e_g$, O $p$, and combined contributions along a high-symmetry path in the BZ from R to $\Gamma$ to X.}\label{fig:kPDOS_SrMnO3}
\end{figure*}

Fig.~\ref{fig:kPDOS_SrMnO3} shows the k-resolved partial DOS along $R-\Gamma-X$. The hybridization between Mn $t_{2g}$ and O $p_{\pi}$ is clearly visible in the k-resolved partial DOS. The indirect gap is formed between Mn $t_{2g}$+ O $p_{\pi}$ valence states at $R$ and Mn $e_{g}$ conduction band states at the $\Gamma$ point, which is consistent with our conclusion from the local DOS in Fig.~\ref{fig:PDOS_SrMnO3} and ~\ref{fig:Mn3d_O2p_vs_Expt}.

Several previous works have studied the electronic structure of the low-temperature AFM SrMnO$_{3}$ using density functional theory (DFT)~\cite{Rune06,SrMnFeO3_expt_Kim10}, and variants of $GW$~\cite{Ergonenc18,GW_EDMFT_PRM_Philipp20}, LDA+DMFT~\cite{Mravlje12} and $GW$+DMFT~\cite{GW_EDMFT_PRM_Philipp20}. 
On the other hand, understanding of the high-temperature PM phase is limited. 
LDA incorrectly predicts a metallic phase for PM SrMnO$_{3}$~\cite{Dang14,Chen14,Bauernfeind18}, suggesting that a mean-field description is insufficient. 
Furthermore, $G_{0}W_{0}$ on top of DFT also fails to open a gap~\cite{GW_EDMFT_PRM_Philipp20}. 
To address the missing correlations, LDA+DMFT can be applied, assuming that strong correlations are purely local on the Mn atom~\cite{Dang14,Chen14,Bauernfeind18}. 
In that case, both the $U$ and $J$ model parameters need to be chosen \emph{ad hoc} in order to reproduce experimental data, and the double counting has to be fine-tuned to keep the system insulating. 
The failure of standard FLL double counting likely comes from the strong hybridization between Mn $3d$ and O $2p$. The band gap, as well as the type of insulating state (e.g. Mott or charge-transfer), are highly sensitive to the parameter choice.

\textcite{GW_EDMFT_PRM_Philipp20} applied multitier GW+EDMFT to both the PM and AFM phase. Even though a gap opening for AFM is observed, a strongly correlated metal in proximity to a Mott transition is observed in PM phase.  
This is likely due to the lack of outer-loop self-consistency for the entire system. 
Such an outer loop is responsible for updating all the weakly correlated orbitals to include strong correlation effects coming form the solution of the impurity problem.
In order to verify this, we have performed SEET calculations without the outer-loop self-consistency for all impurity choices. 
We observe that even though the non-perturbative treatments of the impurities greatly suppresses the DOS at $E_{F}$, the lack of outer-loop self-consistency results in an incorrect chemical potential shift so that a non-zero DOS is observed (see Fig.~\ref{fig:SEET_SrMnO3} in appendix E).
Note that a similar non-zero DOS is observed  in  Ref.~\cite{GW_EDMFT_PRM_Philipp20}. 
The metallic character observed in Ref.~\cite{GW_EDMFT_PRM_Philipp20} may also arise due the missing correlations beyond \emph{GW} from O $2p$ or a failure of cRPA to produce correct screened interactions.

\section{Conclusions}\label{sec:conc}
We have analyzed single-particle spectral functions of two paradigmatic strongly correlated cubic perovskites.
Our starting point was the description of the solid in a  Bloch-wave basis consisting of Gaussian orbitals, which allows for a clear attribution of correlation physics to individual atomic orbitals.
By analyzing strong correlations in multiple choices of impurity orbitals, we showed that the usual procedure of isolating $3$-$5$ local orbitals near the Fermi energy is insufficient to describe the physics of these systems if ab-initio Coulomb interactions are used. 
We have also illustrated that an interpretation in terms of `Hund's physics' with empirically adjusted interaction parameters of low-lying bands is not necessary. 
Rather, by systematically adding non-perturbative correlations, also to orbitals not immediately adjacent to the Fermi energy, we could show a gradual convergence of most aspects of the spectral function to the experimental result.

In the case of SrVO$_3$, we presented that the standard DMFT interpretation of the material as a correlated 3-orbital `Hunds' metal with a three-peak structure of lower and upper Hubbard side-peaks, as well as a central quasiparticle peak, is not consistent with our formulation. 
Through the explicit inclusion of O $2p$ impurities, we also showed that O $2p$ orbitals do not contribute to the lower/upper incoherent feature at around -1.8/3.0 eV. 
The absence of the lower satellite from our calculations indicates that additional physics arising either from non-local processes, higher bands, or/and inter-orbital cross-correlations is important. 
In the case of SrMnO$_3$, we illustrated the importance of the feedback of the strongly correlated self-energy to the remainder of weakly correlated orbitals present in the system, leading successfully to an insulating behavior in PM phase.

In contrast to other methods commonly in use, the procedures employed in this paper are uniquely defined in terms of Gaussian basis sets, choice of correlated orbitals, and choice of correlated subspaces, making every step of the calculation independently reproducible.

Our work illustrates that the SEET procedure is able to recover correct results in strongly-correlated realistic systems, and that systematic addition of correlations in orbital subgroups leads to a systematic understanding of single-particle excitation spectra.

\section{Acknowledgement}
S.I. and E.G. are supported by the Simons foundation via the Simons Collaboration on the Many-Electron Problem, D.Z. and C.-N.Y. by the U.S. Department of Energy under Award No. de-sc0019374. This research used resources of the National Energy Research Scientific Computing Center (NERSC), a U. S. Department of Energy Office of Science User Facility operated under Contract No. DE-AC02-05CH11231.

\bibliographystyle{apsrev4-2}
\bibliography{refs}

\begin{thebibliography}{61}%
\makeatletter
\providecommand \@ifxundefined [1]{%
 \@ifx{#1\undefined}
}%
\providecommand \@ifnum [1]{%
 \ifnum #1\expandafter \@firstoftwo
 \else \expandafter \@secondoftwo
 \fi
}%
\providecommand \@ifx [1]{%
 \ifx #1\expandafter \@firstoftwo
 \else \expandafter \@secondoftwo
 \fi
}%
\providecommand \natexlab [1]{#1}%
\providecommand \enquote  [1]{``#1''}%
\providecommand \bibnamefont  [1]{#1}%
\providecommand \bibfnamefont [1]{#1}%
\providecommand \citenamefont [1]{#1}%
\providecommand \href@noop [0]{\@secondoftwo}%
\providecommand \href [0]{\begingroup \@sanitize@url \@href}%
\providecommand \@href[1]{\@@startlink{#1}\@@href}%
\providecommand \@@href[1]{\endgroup#1\@@endlink}%
\providecommand \@sanitize@url [0]{\catcode `\\12\catcode `\$12\catcode
  `\&12\catcode `\#12\catcode `\^12\catcode `\_12\catcode `\%12\relax}%
\providecommand \@@startlink[1]{}%
\providecommand \@@endlink[0]{}%
\providecommand \url  [0]{\begingroup\@sanitize@url \@url }%
\providecommand \@url [1]{\endgroup\@href {#1}{\urlprefix }}%
\providecommand \urlprefix  [0]{URL }%
\providecommand \Eprint [0]{\href }%
\providecommand \doibase [0]{https://doi.org/}%
\providecommand \selectlanguage [0]{\@gobble}%
\providecommand \bibinfo  [0]{\@secondoftwo}%
\providecommand \bibfield  [0]{\@secondoftwo}%
\providecommand \translation [1]{[#1]}%
\providecommand \BibitemOpen [0]{}%
\providecommand \bibitemStop [0]{}%
\providecommand \bibitemNoStop [0]{.\EOS\space}%
\providecommand \EOS [0]{\spacefactor3000\relax}%
\providecommand \BibitemShut  [1]{\csname bibitem#1\endcsname}%
\let\auto@bib@innerbib\@empty
%</preamble>
\bibitem [{\citenamefont {Pavarini}\ \emph {et~al.}(2004)\citenamefont
  {Pavarini}, \citenamefont {Biermann}, \citenamefont {Poteryaev},
  \citenamefont {Lichtenstein}, \citenamefont {Georges},\ and\ \citenamefont
  {Andersen}}]{Pavarini04}%
  \BibitemOpen
  \bibfield  {author} {\bibinfo {author} {\bibfnamefont {E.}~\bibnamefont
  {Pavarini}}, \bibinfo {author} {\bibfnamefont {S.}~\bibnamefont {Biermann}},
  \bibinfo {author} {\bibfnamefont {A.}~\bibnamefont {Poteryaev}}, \bibinfo
  {author} {\bibfnamefont {A.~I.}\ \bibnamefont {Lichtenstein}}, \bibinfo
  {author} {\bibfnamefont {A.}~\bibnamefont {Georges}},\ and\ \bibinfo {author}
  {\bibfnamefont {O.~K.}\ \bibnamefont {Andersen}},\ }\href
  {https://doi.org/10.1103/PhysRevLett.92.176403} {\bibfield  {journal}
  {\bibinfo  {journal} {Phys. Rev. Lett.}\ }\textbf {\bibinfo {volume} {92}},\
  \bibinfo {pages} {176403} (\bibinfo {year} {2004})}\BibitemShut {NoStop}%
\bibitem [{\citenamefont {Sekiyama}\ \emph {et~al.}(2004)\citenamefont
  {Sekiyama}, \citenamefont {Fujiwara}, \citenamefont {Imada}, \citenamefont
  {Suga}, \citenamefont {Eisaki}, \citenamefont {Uchida}, \citenamefont
  {Takegahara}, \citenamefont {Harima}, \citenamefont {Saitoh}, \citenamefont
  {Nekrasov}, \citenamefont {Keller}, \citenamefont {Kondakov}, \citenamefont
  {Kozhevnikov}, \citenamefont {Pruschke}, \citenamefont {Held}, \citenamefont
  {Vollhardt},\ and\ \citenamefont {Anisimov}}]{SrVO3_Sekiyama04}%
  \BibitemOpen
  \bibfield  {author} {\bibinfo {author} {\bibfnamefont {A.}~\bibnamefont
  {Sekiyama}}, \bibinfo {author} {\bibfnamefont {H.}~\bibnamefont {Fujiwara}},
  \bibinfo {author} {\bibfnamefont {S.}~\bibnamefont {Imada}}, \bibinfo
  {author} {\bibfnamefont {S.}~\bibnamefont {Suga}}, \bibinfo {author}
  {\bibfnamefont {H.}~\bibnamefont {Eisaki}}, \bibinfo {author} {\bibfnamefont
  {S.~I.}\ \bibnamefont {Uchida}}, \bibinfo {author} {\bibfnamefont
  {K.}~\bibnamefont {Takegahara}}, \bibinfo {author} {\bibfnamefont
  {H.}~\bibnamefont {Harima}}, \bibinfo {author} {\bibfnamefont
  {Y.}~\bibnamefont {Saitoh}}, \bibinfo {author} {\bibfnamefont {I.~A.}\
  \bibnamefont {Nekrasov}}, \bibinfo {author} {\bibfnamefont {G.}~\bibnamefont
  {Keller}}, \bibinfo {author} {\bibfnamefont {D.~E.}\ \bibnamefont
  {Kondakov}}, \bibinfo {author} {\bibfnamefont {A.~V.}\ \bibnamefont
  {Kozhevnikov}}, \bibinfo {author} {\bibfnamefont {T.}~\bibnamefont
  {Pruschke}}, \bibinfo {author} {\bibfnamefont {K.}~\bibnamefont {Held}},
  \bibinfo {author} {\bibfnamefont {D.}~\bibnamefont {Vollhardt}},\ and\
  \bibinfo {author} {\bibfnamefont {V.~I.}\ \bibnamefont {Anisimov}},\ }\href
  {https://doi.org/10.1103/PhysRevLett.93.156402} {\bibfield  {journal}
  {\bibinfo  {journal} {Phys. Rev. Lett.}\ }\textbf {\bibinfo {volume} {93}},\
  \bibinfo {pages} {156402} (\bibinfo {year} {2004})}\BibitemShut {NoStop}%
\bibitem [{\citenamefont {Lechermann}\ \emph {et~al.}(2006)\citenamefont
  {Lechermann}, \citenamefont {Georges}, \citenamefont {Poteryaev},
  \citenamefont {Biermann}, \citenamefont {Posternak}, \citenamefont
  {Yamasaki},\ and\ \citenamefont {Andersen}}]{Lechermann06}%
  \BibitemOpen
  \bibfield  {author} {\bibinfo {author} {\bibfnamefont {F.}~\bibnamefont
  {Lechermann}}, \bibinfo {author} {\bibfnamefont {A.}~\bibnamefont {Georges}},
  \bibinfo {author} {\bibfnamefont {A.}~\bibnamefont {Poteryaev}}, \bibinfo
  {author} {\bibfnamefont {S.}~\bibnamefont {Biermann}}, \bibinfo {author}
  {\bibfnamefont {M.}~\bibnamefont {Posternak}}, \bibinfo {author}
  {\bibfnamefont {A.}~\bibnamefont {Yamasaki}},\ and\ \bibinfo {author}
  {\bibfnamefont {O.~K.}\ \bibnamefont {Andersen}},\ }\href
  {https://doi.org/10.1103/PhysRevB.74.125120} {\bibfield  {journal} {\bibinfo
  {journal} {Phys. Rev. B}\ }\textbf {\bibinfo {volume} {74}},\ \bibinfo
  {pages} {125120} (\bibinfo {year} {2006})}\BibitemShut {NoStop}%
\bibitem [{\citenamefont {Nekrasov}\ \emph {et~al.}(2005)\citenamefont
  {Nekrasov}, \citenamefont {Keller}, \citenamefont {Kondakov}, \citenamefont
  {Kozhevnikov}, \citenamefont {Pruschke}, \citenamefont {Held}, \citenamefont
  {Vollhardt},\ and\ \citenamefont {Anisimov}}]{SrVO3_Nekrasov05}%
  \BibitemOpen
  \bibfield  {author} {\bibinfo {author} {\bibfnamefont {I.~A.}\ \bibnamefont
  {Nekrasov}}, \bibinfo {author} {\bibfnamefont {G.}~\bibnamefont {Keller}},
  \bibinfo {author} {\bibfnamefont {D.~E.}\ \bibnamefont {Kondakov}}, \bibinfo
  {author} {\bibfnamefont {A.~V.}\ \bibnamefont {Kozhevnikov}}, \bibinfo
  {author} {\bibfnamefont {T.}~\bibnamefont {Pruschke}}, \bibinfo {author}
  {\bibfnamefont {K.}~\bibnamefont {Held}}, \bibinfo {author} {\bibfnamefont
  {D.}~\bibnamefont {Vollhardt}},\ and\ \bibinfo {author} {\bibfnamefont
  {V.~I.}\ \bibnamefont {Anisimov}},\ }\href
  {https://doi.org/10.1103/PhysRevB.72.155106} {\bibfield  {journal} {\bibinfo
  {journal} {Phys. Rev. B}\ }\textbf {\bibinfo {volume} {72}},\ \bibinfo
  {pages} {155106} (\bibinfo {year} {2005})}\BibitemShut {NoStop}%
\bibitem [{\citenamefont {Nekrasov}\ \emph {et~al.}(2006)\citenamefont
  {Nekrasov}, \citenamefont {Held}, \citenamefont {Keller}, \citenamefont
  {Kondakov}, \citenamefont {Pruschke}, \citenamefont {Kollar}, \citenamefont
  {Andersen}, \citenamefont {Anisimov},\ and\ \citenamefont
  {Vollhardt}}]{Nekrasov06}%
  \BibitemOpen
  \bibfield  {author} {\bibinfo {author} {\bibfnamefont {I.~A.}\ \bibnamefont
  {Nekrasov}}, \bibinfo {author} {\bibfnamefont {K.}~\bibnamefont {Held}},
  \bibinfo {author} {\bibfnamefont {G.}~\bibnamefont {Keller}}, \bibinfo
  {author} {\bibfnamefont {D.~E.}\ \bibnamefont {Kondakov}}, \bibinfo {author}
  {\bibfnamefont {T.}~\bibnamefont {Pruschke}}, \bibinfo {author}
  {\bibfnamefont {M.}~\bibnamefont {Kollar}}, \bibinfo {author} {\bibfnamefont
  {O.~K.}\ \bibnamefont {Andersen}}, \bibinfo {author} {\bibfnamefont {V.~I.}\
  \bibnamefont {Anisimov}},\ and\ \bibinfo {author} {\bibfnamefont
  {D.}~\bibnamefont {Vollhardt}},\ }\href
  {https://doi.org/10.1103/PhysRevB.73.155112} {\bibfield  {journal} {\bibinfo
  {journal} {Phys. Rev. B}\ }\textbf {\bibinfo {volume} {73}},\ \bibinfo
  {pages} {155112} (\bibinfo {year} {2006})}\BibitemShut {NoStop}%
\bibitem [{\citenamefont {Taranto}\ \emph {et~al.}(2013)\citenamefont
  {Taranto}, \citenamefont {Kaltak}, \citenamefont {Parragh}, \citenamefont
  {Sangiovanni}, \citenamefont {Kresse}, \citenamefont {Toschi},\ and\
  \citenamefont {Held}}]{SrVO3_Taranto13}%
  \BibitemOpen
  \bibfield  {author} {\bibinfo {author} {\bibfnamefont {C.}~\bibnamefont
  {Taranto}}, \bibinfo {author} {\bibfnamefont {M.}~\bibnamefont {Kaltak}},
  \bibinfo {author} {\bibfnamefont {N.}~\bibnamefont {Parragh}}, \bibinfo
  {author} {\bibfnamefont {G.}~\bibnamefont {Sangiovanni}}, \bibinfo {author}
  {\bibfnamefont {G.}~\bibnamefont {Kresse}}, \bibinfo {author} {\bibfnamefont
  {A.}~\bibnamefont {Toschi}},\ and\ \bibinfo {author} {\bibfnamefont
  {K.}~\bibnamefont {Held}},\ }\href
  {https://doi.org/10.1103/PhysRevB.88.165119} {\bibfield  {journal} {\bibinfo
  {journal} {Phys. Rev. B}\ }\textbf {\bibinfo {volume} {88}},\ \bibinfo
  {pages} {165119} (\bibinfo {year} {2013})}\BibitemShut {NoStop}%
\bibitem [{\citenamefont {Tomczak}\ \emph {et~al.}(2012)\citenamefont
  {Tomczak}, \citenamefont {Casula}, \citenamefont {Miyake}, \citenamefont
  {Aryasetiawan},\ and\ \citenamefont {Biermann}}]{SrVO3_Tomczak12}%
  \BibitemOpen
  \bibfield  {author} {\bibinfo {author} {\bibfnamefont {J.~M.}\ \bibnamefont
  {Tomczak}}, \bibinfo {author} {\bibfnamefont {M.}~\bibnamefont {Casula}},
  \bibinfo {author} {\bibfnamefont {T.}~\bibnamefont {Miyake}}, \bibinfo
  {author} {\bibfnamefont {F.}~\bibnamefont {Aryasetiawan}},\ and\ \bibinfo
  {author} {\bibfnamefont {S.}~\bibnamefont {Biermann}},\ }\href
  {https://doi.org/10.1209/0295-5075/100/67001} {\bibfield  {journal} {\bibinfo
   {journal} {{EPL} (Europhysics Letters)}\ }\textbf {\bibinfo {volume}
  {100}},\ \bibinfo {pages} {67001} (\bibinfo {year} {2012})}\BibitemShut
  {NoStop}%
\bibitem [{\citenamefont {Sakuma}\ \emph {et~al.}(2013)\citenamefont {Sakuma},
  \citenamefont {Werner},\ and\ \citenamefont {Aryasetiawan}}]{SrVO3_Sakuma13}%
  \BibitemOpen
  \bibfield  {author} {\bibinfo {author} {\bibfnamefont {R.}~\bibnamefont
  {Sakuma}}, \bibinfo {author} {\bibfnamefont {P.}~\bibnamefont {Werner}},\
  and\ \bibinfo {author} {\bibfnamefont {F.}~\bibnamefont {Aryasetiawan}},\
  }\href {https://doi.org/10.1103/PhysRevB.88.235110} {\bibfield  {journal}
  {\bibinfo  {journal} {Phys. Rev. B}\ }\textbf {\bibinfo {volume} {88}},\
  \bibinfo {pages} {235110} (\bibinfo {year} {2013})}\BibitemShut {NoStop}%
\bibitem [{\citenamefont {Dang}\ \emph {et~al.}(2014)\citenamefont {Dang},
  \citenamefont {Ai}, \citenamefont {Millis},\ and\ \citenamefont
  {Marianetti}}]{Dang14}%
  \BibitemOpen
  \bibfield  {author} {\bibinfo {author} {\bibfnamefont {H.~T.}\ \bibnamefont
  {Dang}}, \bibinfo {author} {\bibfnamefont {X.}~\bibnamefont {Ai}}, \bibinfo
  {author} {\bibfnamefont {A.~J.}\ \bibnamefont {Millis}},\ and\ \bibinfo
  {author} {\bibfnamefont {C.~A.}\ \bibnamefont {Marianetti}},\ }\href
  {https://doi.org/10.1103/PhysRevB.90.125114} {\bibfield  {journal} {\bibinfo
  {journal} {Phys. Rev. B}\ }\textbf {\bibinfo {volume} {90}},\ \bibinfo
  {pages} {125114} (\bibinfo {year} {2014})}\BibitemShut {NoStop}%
\bibitem [{\citenamefont {Chen}\ \emph {et~al.}(2014)\citenamefont {Chen},
  \citenamefont {Park}, \citenamefont {Millis},\ and\ \citenamefont
  {Marianetti}}]{Chen14}%
  \BibitemOpen
  \bibfield  {author} {\bibinfo {author} {\bibfnamefont {H.}~\bibnamefont
  {Chen}}, \bibinfo {author} {\bibfnamefont {H.}~\bibnamefont {Park}}, \bibinfo
  {author} {\bibfnamefont {A.~J.}\ \bibnamefont {Millis}},\ and\ \bibinfo
  {author} {\bibfnamefont {C.~A.}\ \bibnamefont {Marianetti}},\ }\href
  {https://doi.org/10.1103/PhysRevB.90.245138} {\bibfield  {journal} {\bibinfo
  {journal} {Phys. Rev. B}\ }\textbf {\bibinfo {volume} {90}},\ \bibinfo
  {pages} {245138} (\bibinfo {year} {2014})}\BibitemShut {NoStop}%
\bibitem [{\citenamefont {Tomczak}\ \emph {et~al.}(2014)\citenamefont
  {Tomczak}, \citenamefont {Casula}, \citenamefont {Miyake},\ and\
  \citenamefont {Biermann}}]{Tomczak14}%
  \BibitemOpen
  \bibfield  {author} {\bibinfo {author} {\bibfnamefont {J.~M.}\ \bibnamefont
  {Tomczak}}, \bibinfo {author} {\bibfnamefont {M.}~\bibnamefont {Casula}},
  \bibinfo {author} {\bibfnamefont {T.}~\bibnamefont {Miyake}},\ and\ \bibinfo
  {author} {\bibfnamefont {S.}~\bibnamefont {Biermann}},\ }\href
  {https://doi.org/10.1103/PhysRevB.90.165138} {\bibfield  {journal} {\bibinfo
  {journal} {Phys. Rev. B}\ }\textbf {\bibinfo {volume} {90}},\ \bibinfo
  {pages} {165138} (\bibinfo {year} {2014})}\BibitemShut {NoStop}%
\bibitem [{\citenamefont {Bauernfeind}\ \emph {et~al.}(2017)\citenamefont
  {Bauernfeind}, \citenamefont {Zingl}, \citenamefont {Triebl}, \citenamefont
  {Aichhorn},\ and\ \citenamefont {Evertz}}]{Bauernfeind17}%
  \BibitemOpen
  \bibfield  {author} {\bibinfo {author} {\bibfnamefont {D.}~\bibnamefont
  {Bauernfeind}}, \bibinfo {author} {\bibfnamefont {M.}~\bibnamefont {Zingl}},
  \bibinfo {author} {\bibfnamefont {R.}~\bibnamefont {Triebl}}, \bibinfo
  {author} {\bibfnamefont {M.}~\bibnamefont {Aichhorn}},\ and\ \bibinfo
  {author} {\bibfnamefont {H.~G.}\ \bibnamefont {Evertz}},\ }\href
  {https://doi.org/10.1103/PhysRevX.7.031013} {\bibfield  {journal} {\bibinfo
  {journal} {Phys. Rev. X}\ }\textbf {\bibinfo {volume} {7}},\ \bibinfo {pages}
  {031013} (\bibinfo {year} {2017})}\BibitemShut {NoStop}%
\bibitem [{\citenamefont {Bauernfeind}\ \emph {et~al.}(2018)\citenamefont
  {Bauernfeind}, \citenamefont {Triebl}, \citenamefont {Zingl}, \citenamefont
  {Aichhorn},\ and\ \citenamefont {Evertz}}]{Bauernfeind18}%
  \BibitemOpen
  \bibfield  {author} {\bibinfo {author} {\bibfnamefont {D.}~\bibnamefont
  {Bauernfeind}}, \bibinfo {author} {\bibfnamefont {R.}~\bibnamefont {Triebl}},
  \bibinfo {author} {\bibfnamefont {M.}~\bibnamefont {Zingl}}, \bibinfo
  {author} {\bibfnamefont {M.}~\bibnamefont {Aichhorn}},\ and\ \bibinfo
  {author} {\bibfnamefont {H.~G.}\ \bibnamefont {Evertz}},\ }\href
  {https://doi.org/10.1103/PhysRevB.97.115156} {\bibfield  {journal} {\bibinfo
  {journal} {Phys. Rev. B}\ }\textbf {\bibinfo {volume} {97}},\ \bibinfo
  {pages} {115156} (\bibinfo {year} {2018})}\BibitemShut {NoStop}%
\bibitem [{\citenamefont {Gunnarsson}\ \emph {et~al.}(1989)\citenamefont
  {Gunnarsson}, \citenamefont {Andersen}, \citenamefont {Jepsen},\ and\
  \citenamefont {Zaanen}}]{Gunnarsson89}%
  \BibitemOpen
  \bibfield  {author} {\bibinfo {author} {\bibfnamefont {O.}~\bibnamefont
  {Gunnarsson}}, \bibinfo {author} {\bibfnamefont {O.~K.}\ \bibnamefont
  {Andersen}}, \bibinfo {author} {\bibfnamefont {O.}~\bibnamefont {Jepsen}},\
  and\ \bibinfo {author} {\bibfnamefont {J.}~\bibnamefont {Zaanen}},\ }\href
  {https://doi.org/10.1103/PhysRevB.39.1708} {\bibfield  {journal} {\bibinfo
  {journal} {Phys. Rev. B}\ }\textbf {\bibinfo {volume} {39}},\ \bibinfo
  {pages} {1708} (\bibinfo {year} {1989})}\BibitemShut {NoStop}%
\bibitem [{\citenamefont {Boehnke}\ \emph {et~al.}(2016)\citenamefont
  {Boehnke}, \citenamefont {Nilsson}, \citenamefont {Aryasetiawan},\ and\
  \citenamefont {Werner}}]{Boehnke16}%
  \BibitemOpen
  \bibfield  {author} {\bibinfo {author} {\bibfnamefont {L.}~\bibnamefont
  {Boehnke}}, \bibinfo {author} {\bibfnamefont {F.}~\bibnamefont {Nilsson}},
  \bibinfo {author} {\bibfnamefont {F.}~\bibnamefont {Aryasetiawan}},\ and\
  \bibinfo {author} {\bibfnamefont {P.}~\bibnamefont {Werner}},\ }\href
  {https://doi.org/10.1103/PhysRevB.94.201106} {\bibfield  {journal} {\bibinfo
  {journal} {Phys. Rev. B}\ }\textbf {\bibinfo {volume} {94}},\ \bibinfo
  {pages} {201106} (\bibinfo {year} {2016})}\BibitemShut {NoStop}%
\bibitem [{\citenamefont {Nilsson}\ \emph {et~al.}(2017)\citenamefont
  {Nilsson}, \citenamefont {Boehnke}, \citenamefont {Werner},\ and\
  \citenamefont {Aryasetiawan}}]{Nilsson17}%
  \BibitemOpen
  \bibfield  {author} {\bibinfo {author} {\bibfnamefont {F.}~\bibnamefont
  {Nilsson}}, \bibinfo {author} {\bibfnamefont {L.}~\bibnamefont {Boehnke}},
  \bibinfo {author} {\bibfnamefont {P.}~\bibnamefont {Werner}},\ and\ \bibinfo
  {author} {\bibfnamefont {F.}~\bibnamefont {Aryasetiawan}},\ }\href
  {https://doi.org/10.1103/PhysRevMaterials.1.043803} {\bibfield  {journal}
  {\bibinfo  {journal} {Phys. Rev. Materials}\ }\textbf {\bibinfo {volume}
  {1}},\ \bibinfo {pages} {043803} (\bibinfo {year} {2017})}\BibitemShut
  {NoStop}%
\bibitem [{\citenamefont {Petocchi}\ \emph {et~al.}(2020)\citenamefont
  {Petocchi}, \citenamefont {Nilsson}, \citenamefont {Aryasetiawan},\ and\
  \citenamefont {Werner}}]{GW_EDMFT_PRM_Philipp20}%
  \BibitemOpen
  \bibfield  {author} {\bibinfo {author} {\bibfnamefont {F.}~\bibnamefont
  {Petocchi}}, \bibinfo {author} {\bibfnamefont {F.}~\bibnamefont {Nilsson}},
  \bibinfo {author} {\bibfnamefont {F.}~\bibnamefont {Aryasetiawan}},\ and\
  \bibinfo {author} {\bibfnamefont {P.}~\bibnamefont {Werner}},\ }\href
  {https://doi.org/10.1103/PhysRevResearch.2.013191} {\bibfield  {journal}
  {\bibinfo  {journal} {Phys. Rev. Research}\ }\textbf {\bibinfo {volume}
  {2}},\ \bibinfo {pages} {013191} (\bibinfo {year} {2020})}\BibitemShut
  {NoStop}%
\bibitem [{\citenamefont {Aryasetiawan}\ \emph {et~al.}(2004)\citenamefont
  {Aryasetiawan}, \citenamefont {Imada}, \citenamefont {Georges}, \citenamefont
  {Kotliar}, \citenamefont {Biermann},\ and\ \citenamefont
  {Lichtenstein}}]{Aryasetiawan04}%
  \BibitemOpen
  \bibfield  {author} {\bibinfo {author} {\bibfnamefont {F.}~\bibnamefont
  {Aryasetiawan}}, \bibinfo {author} {\bibfnamefont {M.}~\bibnamefont {Imada}},
  \bibinfo {author} {\bibfnamefont {A.}~\bibnamefont {Georges}}, \bibinfo
  {author} {\bibfnamefont {G.}~\bibnamefont {Kotliar}}, \bibinfo {author}
  {\bibfnamefont {S.}~\bibnamefont {Biermann}},\ and\ \bibinfo {author}
  {\bibfnamefont {A.~I.}\ \bibnamefont {Lichtenstein}},\ }\href
  {https://doi.org/10.1103/PhysRevB.70.195104} {\bibfield  {journal} {\bibinfo
  {journal} {Phys. Rev. B}\ }\textbf {\bibinfo {volume} {70}},\ \bibinfo
  {pages} {195104} (\bibinfo {year} {2004})}\BibitemShut {NoStop}%
\bibitem [{\citenamefont {Luttinger}\ and\ \citenamefont
  {Ward}(1960)}]{Luttinger60}%
  \BibitemOpen
  \bibfield  {author} {\bibinfo {author} {\bibfnamefont {J.~M.}\ \bibnamefont
  {Luttinger}}\ and\ \bibinfo {author} {\bibfnamefont {J.~C.}\ \bibnamefont
  {Ward}},\ }\href {https://doi.org/10.1103/PhysRev.118.1417} {\bibfield
  {journal} {\bibinfo  {journal} {Phys. Rev.}\ }\textbf {\bibinfo {volume}
  {118}},\ \bibinfo {pages} {1417} (\bibinfo {year} {1960})}\BibitemShut
  {NoStop}%
\bibitem [{\citenamefont {Baym}\ and\ \citenamefont {Kadanoff}(1961)}]{Baym61}%
  \BibitemOpen
  \bibfield  {author} {\bibinfo {author} {\bibfnamefont {G.}~\bibnamefont
  {Baym}}\ and\ \bibinfo {author} {\bibfnamefont {L.~P.}\ \bibnamefont
  {Kadanoff}},\ }\href {https://doi.org/10.1103/PhysRev.124.287} {\bibfield
  {journal} {\bibinfo  {journal} {Phys. Rev.}\ }\textbf {\bibinfo {volume}
  {124}},\ \bibinfo {pages} {287} (\bibinfo {year} {1961})}\BibitemShut
  {NoStop}%
\bibitem [{\citenamefont {Baym}(1962)}]{Baym62}%
  \BibitemOpen
  \bibfield  {author} {\bibinfo {author} {\bibfnamefont {G.}~\bibnamefont
  {Baym}},\ }\href {https://doi.org/10.1103/PhysRev.127.1391} {\bibfield
  {journal} {\bibinfo  {journal} {Phys. Rev.}\ }\textbf {\bibinfo {volume}
  {127}},\ \bibinfo {pages} {1391} (\bibinfo {year} {1962})}\BibitemShut
  {NoStop}%
\bibitem [{\citenamefont {Kananenka}\ \emph {et~al.}(2015)\citenamefont
  {Kananenka}, \citenamefont {Gull},\ and\ \citenamefont {Zgid}}]{Kananenka15}%
  \BibitemOpen
  \bibfield  {author} {\bibinfo {author} {\bibfnamefont {A.~A.}\ \bibnamefont
  {Kananenka}}, \bibinfo {author} {\bibfnamefont {E.}~\bibnamefont {Gull}},\
  and\ \bibinfo {author} {\bibfnamefont {D.}~\bibnamefont {Zgid}},\ }\href
  {https://doi.org/10.1103/PhysRevB.91.121111} {\bibfield  {journal} {\bibinfo
  {journal} {Phys. Rev. B}\ }\textbf {\bibinfo {volume} {91}},\ \bibinfo
  {pages} {121111(R)} (\bibinfo {year} {2015})}\BibitemShut {NoStop}%
\bibitem [{\citenamefont {Zgid}\ and\ \citenamefont {Gull}(2017)}]{Zgid17}%
  \BibitemOpen
  \bibfield  {author} {\bibinfo {author} {\bibfnamefont {D.}~\bibnamefont
  {Zgid}}\ and\ \bibinfo {author} {\bibfnamefont {E.}~\bibnamefont {Gull}},\
  }\href {https://doi.org/10.1088/1367-2630/aa5d34} {\bibfield  {journal}
  {\bibinfo  {journal} {New Journal of Physics}\ }\textbf {\bibinfo {volume}
  {19}},\ \bibinfo {pages} {023047} (\bibinfo {year} {2017})}\BibitemShut
  {NoStop}%
\bibitem [{\citenamefont {Rusakov}\ \emph {et~al.}(2019)\citenamefont
  {Rusakov}, \citenamefont {Iskakov}, \citenamefont {Tran},\ and\ \citenamefont
  {Zgid}}]{Rusakov19}%
  \BibitemOpen
  \bibfield  {author} {\bibinfo {author} {\bibfnamefont {A.~A.}\ \bibnamefont
  {Rusakov}}, \bibinfo {author} {\bibfnamefont {S.}~\bibnamefont {Iskakov}},
  \bibinfo {author} {\bibfnamefont {L.~N.}\ \bibnamefont {Tran}},\ and\
  \bibinfo {author} {\bibfnamefont {D.}~\bibnamefont {Zgid}},\ }\href
  {https://doi.org/10.1021/acs.jctc.8b00927} {\bibfield  {journal} {\bibinfo
  {journal} {Journal of Chemical Theory and Computation}\ }\textbf {\bibinfo
  {volume} {15}},\ \bibinfo {pages} {229} (\bibinfo {year} {2019})}\BibitemShut
  {NoStop}%
\bibitem [{\citenamefont {Iskakov}\ \emph {et~al.}(2020)\citenamefont
  {Iskakov}, \citenamefont {Yeh}, \citenamefont {Gull},\ and\ \citenamefont
  {Zgid}}]{Iskakov20}%
  \BibitemOpen
  \bibfield  {author} {\bibinfo {author} {\bibfnamefont {S.}~\bibnamefont
  {Iskakov}}, \bibinfo {author} {\bibfnamefont {C.-N.}\ \bibnamefont {Yeh}},
  \bibinfo {author} {\bibfnamefont {E.}~\bibnamefont {Gull}},\ and\ \bibinfo
  {author} {\bibfnamefont {D.}~\bibnamefont {Zgid}},\ }\href
  {https://doi.org/10.1103/PhysRevB.102.085105} {\bibfield  {journal} {\bibinfo
   {journal} {Phys. Rev. B}\ }\textbf {\bibinfo {volume} {102}},\ \bibinfo
  {pages} {085105} (\bibinfo {year} {2020})}\BibitemShut {NoStop}%
\bibitem [{\citenamefont {Hedin}(1965)}]{Hedin65}%
  \BibitemOpen
  \bibfield  {author} {\bibinfo {author} {\bibfnamefont {L.}~\bibnamefont
  {Hedin}},\ }\href {https://doi.org/10.1103/PhysRev.139.A796} {\bibfield
  {journal} {\bibinfo  {journal} {Phys. Rev.}\ }\textbf {\bibinfo {volume}
  {139}},\ \bibinfo {pages} {A796} (\bibinfo {year} {1965})}\BibitemShut
  {NoStop}%
\bibitem [{\citenamefont {Aryasetiawan}\ and\ \citenamefont
  {Gunnarsson}(1998)}]{Aryasetiawan98}%
  \BibitemOpen
  \bibfield  {author} {\bibinfo {author} {\bibfnamefont {F.}~\bibnamefont
  {Aryasetiawan}}\ and\ \bibinfo {author} {\bibfnamefont {O.}~\bibnamefont
  {Gunnarsson}},\ }\href {https://doi.org/10.1088/0034-4885/61/3/002}
  {\bibfield  {journal} {\bibinfo  {journal} {Reports on Progress in Physics}\
  }\textbf {\bibinfo {volume} {61}},\ \bibinfo {pages} {237} (\bibinfo {year}
  {1998})}\BibitemShut {NoStop}%
\bibitem [{\citenamefont {Kutepov}\ \emph {et~al.}(2009)\citenamefont
  {Kutepov}, \citenamefont {Savrasov},\ and\ \citenamefont
  {Kotliar}}]{Kutepov09}%
  \BibitemOpen
  \bibfield  {author} {\bibinfo {author} {\bibfnamefont {A.}~\bibnamefont
  {Kutepov}}, \bibinfo {author} {\bibfnamefont {S.~Y.}\ \bibnamefont
  {Savrasov}},\ and\ \bibinfo {author} {\bibfnamefont {G.}~\bibnamefont
  {Kotliar}},\ }\href {https://doi.org/10.1103/PhysRevB.80.041103} {\bibfield
  {journal} {\bibinfo  {journal} {Phys. Rev. B}\ }\textbf {\bibinfo {volume}
  {80}},\ \bibinfo {pages} {041103} (\bibinfo {year} {2009})}\BibitemShut
  {NoStop}%
\bibitem [{\citenamefont {Caffarel}\ and\ \citenamefont
  {Krauth}(1994)}]{Caffarel94}%
  \BibitemOpen
  \bibfield  {author} {\bibinfo {author} {\bibfnamefont {M.}~\bibnamefont
  {Caffarel}}\ and\ \bibinfo {author} {\bibfnamefont {W.}~\bibnamefont
  {Krauth}},\ }\href {https://doi.org/10.1103/PhysRevLett.72.1545} {\bibfield
  {journal} {\bibinfo  {journal} {Phys. Rev. Lett.}\ }\textbf {\bibinfo
  {volume} {72}},\ \bibinfo {pages} {1545} (\bibinfo {year}
  {1994})}\BibitemShut {NoStop}%
\bibitem [{\citenamefont {Iskakov}\ and\ \citenamefont
  {Danilov}(2018)}]{ED_Sergei18}%
  \BibitemOpen
  \bibfield  {author} {\bibinfo {author} {\bibfnamefont {S.}~\bibnamefont
  {Iskakov}}\ and\ \bibinfo {author} {\bibfnamefont {M.}~\bibnamefont
  {Danilov}},\ }\href
  {https://doi.org/https://doi.org/10.1016/j.cpc.2017.12.016} {\bibfield
  {journal} {\bibinfo  {journal} {Computer Physics Communications}\ }\textbf
  {\bibinfo {volume} {225}},\ \bibinfo {pages} {128 } (\bibinfo {year}
  {2018})}\BibitemShut {NoStop}%
\bibitem [{\citenamefont {Iskakov}\ \emph {et~al.}(2019)\citenamefont
  {Iskakov}, \citenamefont {Rusakov}, \citenamefont {Zgid},\ and\ \citenamefont
  {Gull}}]{Iskakov19}%
  \BibitemOpen
  \bibfield  {author} {\bibinfo {author} {\bibfnamefont {S.}~\bibnamefont
  {Iskakov}}, \bibinfo {author} {\bibfnamefont {A.~A.}\ \bibnamefont
  {Rusakov}}, \bibinfo {author} {\bibfnamefont {D.}~\bibnamefont {Zgid}},\ and\
  \bibinfo {author} {\bibfnamefont {E.}~\bibnamefont {Gull}},\ }\href
  {https://doi.org/10.1103/PhysRevB.100.085112} {\bibfield  {journal} {\bibinfo
   {journal} {Phys. Rev. B}\ }\textbf {\bibinfo {volume} {100}},\ \bibinfo
  {pages} {085112} (\bibinfo {year} {2019})}\BibitemShut {NoStop}%
\bibitem [{\citenamefont {Löwdin}(1970)}]{Lowdin70_SAO}%
  \BibitemOpen
  \bibfield  {author} {\bibinfo {author} {\bibfnamefont {P.-O.}\ \bibnamefont
  {Löwdin}}\ }(\bibinfo  {publisher} {Academic Press},\ \bibinfo {year}
  {1970})\ pp.\ \bibinfo {pages} {185 -- 199}\BibitemShut {NoStop}%
\bibitem [{\citenamefont {Georges}\ \emph {et~al.}(1996)\citenamefont
  {Georges}, \citenamefont {Kotliar}, \citenamefont {Krauth},\ and\
  \citenamefont {Rozenberg}}]{Georges96}%
  \BibitemOpen
  \bibfield  {author} {\bibinfo {author} {\bibfnamefont {A.}~\bibnamefont
  {Georges}}, \bibinfo {author} {\bibfnamefont {G.}~\bibnamefont {Kotliar}},
  \bibinfo {author} {\bibfnamefont {W.}~\bibnamefont {Krauth}},\ and\ \bibinfo
  {author} {\bibfnamefont {M.~J.}\ \bibnamefont {Rozenberg}},\ }\href
  {https://doi.org/10.1103/RevModPhys.68.13} {\bibfield  {journal} {\bibinfo
  {journal} {Rev. Mod. Phys.}\ }\textbf {\bibinfo {volume} {68}},\ \bibinfo
  {pages} {13} (\bibinfo {year} {1996})}\BibitemShut {NoStop}%
\bibitem [{\citenamefont {Kotliar}\ \emph {et~al.}(2006)\citenamefont
  {Kotliar}, \citenamefont {Savrasov}, \citenamefont {Haule}, \citenamefont
  {Oudovenko}, \citenamefont {Parcollet},\ and\ \citenamefont
  {Marianetti}}]{RMP_Kotliar06}%
  \BibitemOpen
  \bibfield  {author} {\bibinfo {author} {\bibfnamefont {G.}~\bibnamefont
  {Kotliar}}, \bibinfo {author} {\bibfnamefont {S.~Y.}\ \bibnamefont
  {Savrasov}}, \bibinfo {author} {\bibfnamefont {K.}~\bibnamefont {Haule}},
  \bibinfo {author} {\bibfnamefont {V.~S.}\ \bibnamefont {Oudovenko}}, \bibinfo
  {author} {\bibfnamefont {O.}~\bibnamefont {Parcollet}},\ and\ \bibinfo
  {author} {\bibfnamefont {C.~A.}\ \bibnamefont {Marianetti}},\ }\href
  {https://doi.org/10.1103/RevModPhys.78.865} {\bibfield  {journal} {\bibinfo
  {journal} {Rev. Mod. Phys.}\ }\textbf {\bibinfo {volume} {78}},\ \bibinfo
  {pages} {865} (\bibinfo {year} {2006})}\BibitemShut {NoStop}%
\bibitem [{\citenamefont {VandeVondele}\ and\ \citenamefont
  {Hutter}(2007)}]{GTHBasis}%
  \BibitemOpen
  \bibfield  {author} {\bibinfo {author} {\bibfnamefont {J.}~\bibnamefont
  {VandeVondele}}\ and\ \bibinfo {author} {\bibfnamefont {J.}~\bibnamefont
  {Hutter}},\ }\href {https://doi.org/10.1063/1.2770708} {\bibfield  {journal}
  {\bibinfo  {journal} {The Journal of Chemical Physics}\ }\textbf {\bibinfo
  {volume} {127}},\ \bibinfo {pages} {114105} (\bibinfo {year}
  {2007})}\BibitemShut {NoStop}%
\bibitem [{\citenamefont {Goedecker}\ \emph {et~al.}(1996)\citenamefont
  {Goedecker}, \citenamefont {Teter},\ and\ \citenamefont
  {Hutter}}]{GTHPseudo}%
  \BibitemOpen
  \bibfield  {author} {\bibinfo {author} {\bibfnamefont {S.}~\bibnamefont
  {Goedecker}}, \bibinfo {author} {\bibfnamefont {M.}~\bibnamefont {Teter}},\
  and\ \bibinfo {author} {\bibfnamefont {J.}~\bibnamefont {Hutter}},\ }\href
  {https://doi.org/https://doi.org/10.1103/PhysRevB.54.1703} {\bibfield
  {journal} {\bibinfo  {journal} {Phys. Rev. B}\ }\textbf {\bibinfo {volume}
  {54}},\ \bibinfo {pages} {1703} (\bibinfo {year} {1996})}\BibitemShut
  {NoStop}%
\bibitem [{\citenamefont {Hättig}(2005)}]{RI_auxbasis}%
  \BibitemOpen
  \bibfield  {author} {\bibinfo {author} {\bibfnamefont {C.}~\bibnamefont
  {Hättig}},\ }\href {https://doi.org/10.1039/B415208E} {\bibfield  {journal}
  {\bibinfo  {journal} {Phys. Chem. Chem. Phys.}\ }\textbf {\bibinfo {volume}
  {7}},\ \bibinfo {pages} {59} (\bibinfo {year} {2005})}\BibitemShut {NoStop}%
\bibitem [{\citenamefont {Sun}\ \emph {et~al.}(2017)\citenamefont {Sun},
  \citenamefont {Berkelbach}, \citenamefont {Blunt}, \citenamefont {Booth},
  \citenamefont {Guo}, \citenamefont {Li}, \citenamefont {Liu}, \citenamefont
  {McClain}, \citenamefont {Sayfutyarova}, \citenamefont {Sharma},
  \citenamefont {Wouters},\ and\ \citenamefont {Chan}}]{PySCF}%
  \BibitemOpen
  \bibfield  {author} {\bibinfo {author} {\bibfnamefont {Q.}~\bibnamefont
  {Sun}}, \bibinfo {author} {\bibfnamefont {T.~C.}\ \bibnamefont {Berkelbach}},
  \bibinfo {author} {\bibfnamefont {N.~S.}\ \bibnamefont {Blunt}}, \bibinfo
  {author} {\bibfnamefont {G.~H.}\ \bibnamefont {Booth}}, \bibinfo {author}
  {\bibfnamefont {S.}~\bibnamefont {Guo}}, \bibinfo {author} {\bibfnamefont
  {Z.}~\bibnamefont {Li}}, \bibinfo {author} {\bibfnamefont {J.}~\bibnamefont
  {Liu}}, \bibinfo {author} {\bibfnamefont {J.~D.}\ \bibnamefont {McClain}},
  \bibinfo {author} {\bibfnamefont {E.~R.}\ \bibnamefont {Sayfutyarova}},
  \bibinfo {author} {\bibfnamefont {S.}~\bibnamefont {Sharma}}, \bibinfo
  {author} {\bibfnamefont {S.}~\bibnamefont {Wouters}},\ and\ \bibinfo {author}
  {\bibfnamefont {G.~K.}\ \bibnamefont {Chan}},\ }\href
  {https://doi.org/10.1002/wcms.1340} {\bibinfo {title} {Pyscf: the
  python-based simulations of chemistry framework}} (\bibinfo {year}
  {2017})\BibitemShut {NoStop}%
\bibitem [{\citenamefont {Shinaoka}\ \emph {et~al.}(2017)\citenamefont
  {Shinaoka}, \citenamefont {Otsuki}, \citenamefont {Ohzeki},\ and\
  \citenamefont {Yoshimi}}]{Shinaoka17}%
  \BibitemOpen
  \bibfield  {author} {\bibinfo {author} {\bibfnamefont {H.}~\bibnamefont
  {Shinaoka}}, \bibinfo {author} {\bibfnamefont {J.}~\bibnamefont {Otsuki}},
  \bibinfo {author} {\bibfnamefont {M.}~\bibnamefont {Ohzeki}},\ and\ \bibinfo
  {author} {\bibfnamefont {K.}~\bibnamefont {Yoshimi}},\ }\href
  {https://doi.org/10.1103/PhysRevB.96.035147} {\bibfield  {journal} {\bibinfo
  {journal} {Phys. Rev. B}\ }\textbf {\bibinfo {volume} {96}},\ \bibinfo
  {pages} {035147} (\bibinfo {year} {2017})}\BibitemShut {NoStop}%
\bibitem [{\citenamefont {Li}\ \emph {et~al.}(2020)\citenamefont {Li},
  \citenamefont {Wallerberger}, \citenamefont {Chikano}, \citenamefont {Yeh},
  \citenamefont {Gull},\ and\ \citenamefont {Shinaoka}}]{Li20}%
  \BibitemOpen
  \bibfield  {author} {\bibinfo {author} {\bibfnamefont {J.}~\bibnamefont
  {Li}}, \bibinfo {author} {\bibfnamefont {M.}~\bibnamefont {Wallerberger}},
  \bibinfo {author} {\bibfnamefont {N.}~\bibnamefont {Chikano}}, \bibinfo
  {author} {\bibfnamefont {C.-N.}\ \bibnamefont {Yeh}}, \bibinfo {author}
  {\bibfnamefont {E.}~\bibnamefont {Gull}},\ and\ \bibinfo {author}
  {\bibfnamefont {H.}~\bibnamefont {Shinaoka}},\ }\href
  {https://doi.org/10.1103/PhysRevB.101.035144} {\bibfield  {journal} {\bibinfo
   {journal} {Phys. Rev. B}\ }\textbf {\bibinfo {volume} {101}},\ \bibinfo
  {pages} {035144} (\bibinfo {year} {2020})}\BibitemShut {NoStop}%
\bibitem [{\citenamefont {Zhu}\ and\ \citenamefont {Chan}(2020)}]{zhu2020ab}%
  \BibitemOpen
  \bibfield  {author} {\bibinfo {author} {\bibfnamefont {T.}~\bibnamefont
  {Zhu}}\ and\ \bibinfo {author} {\bibfnamefont {G.~K.-L.}\ \bibnamefont
  {Chan}},\ }\href@noop {} {\bibinfo {title} {Ab initio full cell gw+dmft for
  correlated materials}} (\bibinfo {year} {2020}),\ \Eprint
  {https://arxiv.org/abs/2003.01349} {arXiv:2003.01349 [cond-mat.str-el]}
  \BibitemShut {NoStop}%
\bibitem [{\citenamefont {Yoshimatsu}\ \emph {et~al.}(2010)\citenamefont
  {Yoshimatsu}, \citenamefont {Okabe}, \citenamefont {Kumigashira},
  \citenamefont {Okamoto}, \citenamefont {Aizaki}, \citenamefont {Fujimori},\
  and\ \citenamefont {Oshima}}]{SrVO3_PES_Yoshimatsu10}%
  \BibitemOpen
  \bibfield  {author} {\bibinfo {author} {\bibfnamefont {K.}~\bibnamefont
  {Yoshimatsu}}, \bibinfo {author} {\bibfnamefont {T.}~\bibnamefont {Okabe}},
  \bibinfo {author} {\bibfnamefont {H.}~\bibnamefont {Kumigashira}}, \bibinfo
  {author} {\bibfnamefont {S.}~\bibnamefont {Okamoto}}, \bibinfo {author}
  {\bibfnamefont {S.}~\bibnamefont {Aizaki}}, \bibinfo {author} {\bibfnamefont
  {A.}~\bibnamefont {Fujimori}},\ and\ \bibinfo {author} {\bibfnamefont
  {M.}~\bibnamefont {Oshima}},\ }\href
  {https://doi.org/10.1103/PhysRevLett.104.147601} {\bibfield  {journal}
  {\bibinfo  {journal} {Phys. Rev. Lett.}\ }\textbf {\bibinfo {volume} {104}},\
  \bibinfo {pages} {147601} (\bibinfo {year} {2010})}\BibitemShut {NoStop}%
\bibitem [{\citenamefont {Morikawa}\ \emph {et~al.}(1995)\citenamefont
  {Morikawa}, \citenamefont {Mizokawa}, \citenamefont {Kobayashi},
  \citenamefont {Fujimori}, \citenamefont {Eisaki}, \citenamefont {Uchida},
  \citenamefont {Iga},\ and\ \citenamefont {Nishihara}}]{SrVO3_Morikawa95}%
  \BibitemOpen
  \bibfield  {author} {\bibinfo {author} {\bibfnamefont {K.}~\bibnamefont
  {Morikawa}}, \bibinfo {author} {\bibfnamefont {T.}~\bibnamefont {Mizokawa}},
  \bibinfo {author} {\bibfnamefont {K.}~\bibnamefont {Kobayashi}}, \bibinfo
  {author} {\bibfnamefont {A.}~\bibnamefont {Fujimori}}, \bibinfo {author}
  {\bibfnamefont {H.}~\bibnamefont {Eisaki}}, \bibinfo {author} {\bibfnamefont
  {S.}~\bibnamefont {Uchida}}, \bibinfo {author} {\bibfnamefont
  {F.}~\bibnamefont {Iga}},\ and\ \bibinfo {author} {\bibfnamefont
  {Y.}~\bibnamefont {Nishihara}},\ }\href
  {https://doi.org/10.1103/PhysRevB.52.13711} {\bibfield  {journal} {\bibinfo
  {journal} {Phys. Rev. B}\ }\textbf {\bibinfo {volume} {52}},\ \bibinfo
  {pages} {13711} (\bibinfo {year} {1995})}\BibitemShut {NoStop}%
\bibitem [{\citenamefont {Takizawa}\ \emph {et~al.}(2009)\citenamefont
  {Takizawa}, \citenamefont {Minohara}, \citenamefont {Kumigashira},
  \citenamefont {Toyota}, \citenamefont {Oshima}, \citenamefont {Wadati},
  \citenamefont {Yoshida}, \citenamefont {Fujimori}, \citenamefont {Lippmaa},
  \citenamefont {Kawasaki}, \citenamefont {Koinuma}, \citenamefont {Sordi},\
  and\ \citenamefont {Rozenberg}}]{Takizawa09}%
  \BibitemOpen
  \bibfield  {author} {\bibinfo {author} {\bibfnamefont {M.}~\bibnamefont
  {Takizawa}}, \bibinfo {author} {\bibfnamefont {M.}~\bibnamefont {Minohara}},
  \bibinfo {author} {\bibfnamefont {H.}~\bibnamefont {Kumigashira}}, \bibinfo
  {author} {\bibfnamefont {D.}~\bibnamefont {Toyota}}, \bibinfo {author}
  {\bibfnamefont {M.}~\bibnamefont {Oshima}}, \bibinfo {author} {\bibfnamefont
  {H.}~\bibnamefont {Wadati}}, \bibinfo {author} {\bibfnamefont
  {T.}~\bibnamefont {Yoshida}}, \bibinfo {author} {\bibfnamefont
  {A.}~\bibnamefont {Fujimori}}, \bibinfo {author} {\bibfnamefont
  {M.}~\bibnamefont {Lippmaa}}, \bibinfo {author} {\bibfnamefont
  {M.}~\bibnamefont {Kawasaki}}, \bibinfo {author} {\bibfnamefont
  {H.}~\bibnamefont {Koinuma}}, \bibinfo {author} {\bibfnamefont
  {G.}~\bibnamefont {Sordi}},\ and\ \bibinfo {author} {\bibfnamefont
  {M.}~\bibnamefont {Rozenberg}},\ }\href
  {https://doi.org/10.1103/PhysRevB.80.235104} {\bibfield  {journal} {\bibinfo
  {journal} {Phys. Rev. B}\ }\textbf {\bibinfo {volume} {80}},\ \bibinfo
  {pages} {235104} (\bibinfo {year} {2009})}\BibitemShut {NoStop}%
\bibitem [{\citenamefont {Jarrell}\ and\ \citenamefont
  {Gubernatis}(1996)}]{Jarrell96}%
  \BibitemOpen
  \bibfield  {author} {\bibinfo {author} {\bibfnamefont {M.}~\bibnamefont
  {Jarrell}}\ and\ \bibinfo {author} {\bibfnamefont {J.}~\bibnamefont
  {Gubernatis}},\ }\href
  {https://doi.org/https://doi.org/10.1016/0370-1573(95)00074-7} {\bibfield
  {journal} {\bibinfo  {journal} {Physics Reports}\ }\textbf {\bibinfo {volume}
  {269}},\ \bibinfo {pages} {133 } (\bibinfo {year} {1996})}\BibitemShut
  {NoStop}%
\bibitem [{\citenamefont {Levy}\ \emph {et~al.}(2017)\citenamefont {Levy},
  \citenamefont {LeBlanc},\ and\ \citenamefont {Gull}}]{Levy17}%
  \BibitemOpen
  \bibfield  {author} {\bibinfo {author} {\bibfnamefont {R.}~\bibnamefont
  {Levy}}, \bibinfo {author} {\bibfnamefont {J.}~\bibnamefont {LeBlanc}},\ and\
  \bibinfo {author} {\bibfnamefont {E.}~\bibnamefont {Gull}},\ }\href
  {https://doi.org/https://doi.org/10.1016/j.cpc.2017.01.018} {\bibfield
  {journal} {\bibinfo  {journal} {Computer Physics Communications}\ }\textbf
  {\bibinfo {volume} {215}},\ \bibinfo {pages} {149 } (\bibinfo {year}
  {2017})}\BibitemShut {NoStop}%
\bibitem [{\citenamefont {Miyake}\ \emph {et~al.}(2013)\citenamefont {Miyake},
  \citenamefont {Martins}, \citenamefont {Sakuma},\ and\ \citenamefont
  {Aryasetiawan}}]{Miyake13}%
  \BibitemOpen
  \bibfield  {author} {\bibinfo {author} {\bibfnamefont {T.}~\bibnamefont
  {Miyake}}, \bibinfo {author} {\bibfnamefont {C.}~\bibnamefont {Martins}},
  \bibinfo {author} {\bibfnamefont {R.}~\bibnamefont {Sakuma}},\ and\ \bibinfo
  {author} {\bibfnamefont {F.}~\bibnamefont {Aryasetiawan}},\ }\href
  {https://doi.org/10.1103/PhysRevB.87.115110} {\bibfield  {journal} {\bibinfo
  {journal} {Phys. Rev. B}\ }\textbf {\bibinfo {volume} {87}},\ \bibinfo
  {pages} {115110} (\bibinfo {year} {2013})}\BibitemShut {NoStop}%
\bibitem [{\citenamefont {Mizokawa}\ and\ \citenamefont
  {Fujimori}(1996)}]{Mizokawa96}%
  \BibitemOpen
  \bibfield  {author} {\bibinfo {author} {\bibfnamefont {T.}~\bibnamefont
  {Mizokawa}}\ and\ \bibinfo {author} {\bibfnamefont {A.}~\bibnamefont
  {Fujimori}},\ }\href {https://doi.org/10.1103/PhysRevB.54.5368} {\bibfield
  {journal} {\bibinfo  {journal} {Phys. Rev. B}\ }\textbf {\bibinfo {volume}
  {54}},\ \bibinfo {pages} {5368} (\bibinfo {year} {1996})}\BibitemShut
  {NoStop}%
\bibitem [{\citenamefont {Czy\ifmmode~\dot{z}\else \.{z}\fi{}yk}\ and\
  \citenamefont {Sawatzky}(1994)}]{FLL_94}%
  \BibitemOpen
  \bibfield  {author} {\bibinfo {author} {\bibfnamefont {M.~T.}\ \bibnamefont
  {Czy\ifmmode~\dot{z}\else \.{z}\fi{}yk}}\ and\ \bibinfo {author}
  {\bibfnamefont {G.~A.}\ \bibnamefont {Sawatzky}},\ }\href
  {https://doi.org/10.1103/PhysRevB.49.14211} {\bibfield  {journal} {\bibinfo
  {journal} {Phys. Rev. B}\ }\textbf {\bibinfo {volume} {49}},\ \bibinfo
  {pages} {14211} (\bibinfo {year} {1994})}\BibitemShut {NoStop}%
\bibitem [{\citenamefont {Gatti}\ and\ \citenamefont
  {Guzzo}(2013)}]{SrVO3_Gatti13}%
  \BibitemOpen
  \bibfield  {author} {\bibinfo {author} {\bibfnamefont {M.}~\bibnamefont
  {Gatti}}\ and\ \bibinfo {author} {\bibfnamefont {M.}~\bibnamefont {Guzzo}},\
  }\href {https://doi.org/10.1103/PhysRevB.87.155147} {\bibfield  {journal}
  {\bibinfo  {journal} {Phys. Rev. B}\ }\textbf {\bibinfo {volume} {87}},\
  \bibinfo {pages} {155147} (\bibinfo {year} {2013})}\BibitemShut {NoStop}%
\bibitem [{\citenamefont {Yoshida}\ \emph {et~al.}(2005)\citenamefont
  {Yoshida}, \citenamefont {Tanaka}, \citenamefont {Yagi}, \citenamefont {Ino},
  \citenamefont {Eisaki}, \citenamefont {Fujimori},\ and\ \citenamefont
  {Shen}}]{SrVO3_Yoshida05}%
  \BibitemOpen
  \bibfield  {author} {\bibinfo {author} {\bibfnamefont {T.}~\bibnamefont
  {Yoshida}}, \bibinfo {author} {\bibfnamefont {K.}~\bibnamefont {Tanaka}},
  \bibinfo {author} {\bibfnamefont {H.}~\bibnamefont {Yagi}}, \bibinfo {author}
  {\bibfnamefont {A.}~\bibnamefont {Ino}}, \bibinfo {author} {\bibfnamefont
  {H.}~\bibnamefont {Eisaki}}, \bibinfo {author} {\bibfnamefont
  {A.}~\bibnamefont {Fujimori}},\ and\ \bibinfo {author} {\bibfnamefont
  {Z.-X.}\ \bibnamefont {Shen}},\ }\href
  {https://doi.org/10.1103/PhysRevLett.95.146404} {\bibfield  {journal}
  {\bibinfo  {journal} {Phys. Rev. Lett.}\ }\textbf {\bibinfo {volume} {95}},\
  \bibinfo {pages} {146404} (\bibinfo {year} {2005})}\BibitemShut {NoStop}%
\bibitem [{\citenamefont {Kim}\ \emph {et~al.}(2010)\citenamefont {Kim},
  \citenamefont {Lee}, \citenamefont {Dabrowski}, \citenamefont {Kolesnik},
  \citenamefont {Lee}, \citenamefont {Kim}, \citenamefont {Min},\ and\
  \citenamefont {Kang}}]{SrMnFeO3_expt_Kim10}%
  \BibitemOpen
  \bibfield  {author} {\bibinfo {author} {\bibfnamefont {D.~H.}\ \bibnamefont
  {Kim}}, \bibinfo {author} {\bibfnamefont {H.~J.}\ \bibnamefont {Lee}},
  \bibinfo {author} {\bibfnamefont {B.}~\bibnamefont {Dabrowski}}, \bibinfo
  {author} {\bibfnamefont {S.}~\bibnamefont {Kolesnik}}, \bibinfo {author}
  {\bibfnamefont {J.}~\bibnamefont {Lee}}, \bibinfo {author} {\bibfnamefont
  {B.}~\bibnamefont {Kim}}, \bibinfo {author} {\bibfnamefont {B.~I.}\
  \bibnamefont {Min}},\ and\ \bibinfo {author} {\bibfnamefont {J.-S.}\
  \bibnamefont {Kang}},\ }\href {https://doi.org/10.1103/PhysRevB.81.073101}
  {\bibfield  {journal} {\bibinfo  {journal} {Phys. Rev. B}\ }\textbf {\bibinfo
  {volume} {81}},\ \bibinfo {pages} {073101} (\bibinfo {year}
  {2010})}\BibitemShut {NoStop}%
\bibitem [{\citenamefont {Negas}\ and\ \citenamefont {Roth}(1970)}]{Negas70}%
  \BibitemOpen
  \bibfield  {author} {\bibinfo {author} {\bibfnamefont {T.}~\bibnamefont
  {Negas}}\ and\ \bibinfo {author} {\bibfnamefont {R.~S.}\ \bibnamefont
  {Roth}},\ }\href
  {https://doi.org/https://doi.org/10.1016/0022-4596(70)90123-4} {\bibfield
  {journal} {\bibinfo  {journal} {Journal of Solid State Chemistry}\ }\textbf
  {\bibinfo {volume} {1}},\ \bibinfo {pages} {409 } (\bibinfo {year}
  {1970})}\BibitemShut {NoStop}%
\bibitem [{\citenamefont {Takeda}\ and\ \citenamefont
  {Ōhara}(1974)}]{Takeda74}%
  \BibitemOpen
  \bibfield  {author} {\bibinfo {author} {\bibfnamefont {T.}~\bibnamefont
  {Takeda}}\ and\ \bibinfo {author} {\bibfnamefont {S.}~\bibnamefont
  {Ōhara}},\ }\href {https://doi.org/10.1143/JPSJ.37.275} {\bibfield
  {journal} {\bibinfo  {journal} {Journal of the Physical Society of Japan}\
  }\textbf {\bibinfo {volume} {37}},\ \bibinfo {pages} {275} (\bibinfo {year}
  {1974})}\BibitemShut {NoStop}%
\bibitem [{\citenamefont {Abbate}\ \emph {et~al.}(1992)\citenamefont {Abbate},
  \citenamefont {de~Groot}, \citenamefont {Fuggle}, \citenamefont {Fujimori},
  \citenamefont {Strebel}, \citenamefont {Lopez}, \citenamefont {Domke},
  \citenamefont {Kaindl}, \citenamefont {Sawatzky}, \citenamefont {Takano},
  \citenamefont {Takeda}, \citenamefont {Eisaki},\ and\ \citenamefont
  {Uchida}}]{Abbate92}%
  \BibitemOpen
  \bibfield  {author} {\bibinfo {author} {\bibfnamefont {M.}~\bibnamefont
  {Abbate}}, \bibinfo {author} {\bibfnamefont {F.~M.~F.}\ \bibnamefont
  {de~Groot}}, \bibinfo {author} {\bibfnamefont {J.~C.}\ \bibnamefont
  {Fuggle}}, \bibinfo {author} {\bibfnamefont {A.}~\bibnamefont {Fujimori}},
  \bibinfo {author} {\bibfnamefont {O.}~\bibnamefont {Strebel}}, \bibinfo
  {author} {\bibfnamefont {F.}~\bibnamefont {Lopez}}, \bibinfo {author}
  {\bibfnamefont {M.}~\bibnamefont {Domke}}, \bibinfo {author} {\bibfnamefont
  {G.}~\bibnamefont {Kaindl}}, \bibinfo {author} {\bibfnamefont {G.~A.}\
  \bibnamefont {Sawatzky}}, \bibinfo {author} {\bibfnamefont {M.}~\bibnamefont
  {Takano}}, \bibinfo {author} {\bibfnamefont {Y.}~\bibnamefont {Takeda}},
  \bibinfo {author} {\bibfnamefont {H.}~\bibnamefont {Eisaki}},\ and\ \bibinfo
  {author} {\bibfnamefont {S.}~\bibnamefont {Uchida}},\ }\href
  {https://doi.org/10.1103/PhysRevB.46.4511} {\bibfield  {journal} {\bibinfo
  {journal} {Phys. Rev. B}\ }\textbf {\bibinfo {volume} {46}},\ \bibinfo
  {pages} {4511} (\bibinfo {year} {1992})}\BibitemShut {NoStop}%
\bibitem [{\citenamefont {Saitoh}\ \emph {et~al.}(1995)\citenamefont {Saitoh},
  \citenamefont {Bocquet}, \citenamefont {Mizokawa}, \citenamefont {Namatame},
  \citenamefont {Fujimori}, \citenamefont {Abbate}, \citenamefont {Takeda},\
  and\ \citenamefont {Takano}}]{Saitoh95}%
  \BibitemOpen
  \bibfield  {author} {\bibinfo {author} {\bibfnamefont {T.}~\bibnamefont
  {Saitoh}}, \bibinfo {author} {\bibfnamefont {A.~E.}\ \bibnamefont {Bocquet}},
  \bibinfo {author} {\bibfnamefont {T.}~\bibnamefont {Mizokawa}}, \bibinfo
  {author} {\bibfnamefont {H.}~\bibnamefont {Namatame}}, \bibinfo {author}
  {\bibfnamefont {A.}~\bibnamefont {Fujimori}}, \bibinfo {author}
  {\bibfnamefont {M.}~\bibnamefont {Abbate}}, \bibinfo {author} {\bibfnamefont
  {Y.}~\bibnamefont {Takeda}},\ and\ \bibinfo {author} {\bibfnamefont
  {M.}~\bibnamefont {Takano}},\ }\href
  {https://doi.org/10.1103/PhysRevB.51.13942} {\bibfield  {journal} {\bibinfo
  {journal} {Phys. Rev. B}\ }\textbf {\bibinfo {volume} {51}},\ \bibinfo
  {pages} {13942} (\bibinfo {year} {1995})}\BibitemShut {NoStop}%
\bibitem [{\citenamefont {Kang}\ \emph {et~al.}(2008)\citenamefont {Kang},
  \citenamefont {Lee}, \citenamefont {Kim}, \citenamefont {Kim}, \citenamefont
  {Dabrowski}, \citenamefont {Kolesnik}, \citenamefont {Lee}, \citenamefont
  {Kim},\ and\ \citenamefont {Min}}]{Kang08}%
  \BibitemOpen
  \bibfield  {author} {\bibinfo {author} {\bibfnamefont {J.-S.}\ \bibnamefont
  {Kang}}, \bibinfo {author} {\bibfnamefont {H.~J.}\ \bibnamefont {Lee}},
  \bibinfo {author} {\bibfnamefont {G.}~\bibnamefont {Kim}}, \bibinfo {author}
  {\bibfnamefont {D.~H.}\ \bibnamefont {Kim}}, \bibinfo {author} {\bibfnamefont
  {B.}~\bibnamefont {Dabrowski}}, \bibinfo {author} {\bibfnamefont
  {S.}~\bibnamefont {Kolesnik}}, \bibinfo {author} {\bibfnamefont
  {H.}~\bibnamefont {Lee}}, \bibinfo {author} {\bibfnamefont {J.-Y.}\
  \bibnamefont {Kim}},\ and\ \bibinfo {author} {\bibfnamefont {B.~I.}\
  \bibnamefont {Min}},\ }\href {https://doi.org/10.1103/PhysRevB.78.054434}
  {\bibfield  {journal} {\bibinfo  {journal} {Phys. Rev. B}\ }\textbf {\bibinfo
  {volume} {78}},\ \bibinfo {pages} {054434} (\bibinfo {year}
  {2008})}\BibitemShut {NoStop}%
\bibitem [{\citenamefont {S\o{}nden\aa{}}\ \emph {et~al.}(2006)\citenamefont
  {S\o{}nden\aa{}}, \citenamefont {Ravindran}, \citenamefont {St\o{}len},
  \citenamefont {Grande},\ and\ \citenamefont {Hanfland}}]{Rune06}%
  \BibitemOpen
  \bibfield  {author} {\bibinfo {author} {\bibfnamefont {R.}~\bibnamefont
  {S\o{}nden\aa{}}}, \bibinfo {author} {\bibfnamefont {P.}~\bibnamefont
  {Ravindran}}, \bibinfo {author} {\bibfnamefont {S.}~\bibnamefont
  {St\o{}len}}, \bibinfo {author} {\bibfnamefont {T.}~\bibnamefont {Grande}},\
  and\ \bibinfo {author} {\bibfnamefont {M.}~\bibnamefont {Hanfland}},\ }\href
  {https://doi.org/10.1103/PhysRevB.74.144102} {\bibfield  {journal} {\bibinfo
  {journal} {Phys. Rev. B}\ }\textbf {\bibinfo {volume} {74}},\ \bibinfo
  {pages} {144102} (\bibinfo {year} {2006})}\BibitemShut {NoStop}%
\bibitem [{\citenamefont {Erg\"onenc}\ \emph {et~al.}(2018)\citenamefont
  {Erg\"onenc}, \citenamefont {Kim}, \citenamefont {Liu}, \citenamefont
  {Kresse},\ and\ \citenamefont {Franchini}}]{Ergonenc18}%
  \BibitemOpen
  \bibfield  {author} {\bibinfo {author} {\bibfnamefont {Z.}~\bibnamefont
  {Erg\"onenc}}, \bibinfo {author} {\bibfnamefont {B.}~\bibnamefont {Kim}},
  \bibinfo {author} {\bibfnamefont {P.}~\bibnamefont {Liu}}, \bibinfo {author}
  {\bibfnamefont {G.}~\bibnamefont {Kresse}},\ and\ \bibinfo {author}
  {\bibfnamefont {C.}~\bibnamefont {Franchini}},\ }\href
  {https://doi.org/10.1103/PhysRevMaterials.2.024601} {\bibfield  {journal}
  {\bibinfo  {journal} {Phys. Rev. Materials}\ }\textbf {\bibinfo {volume}
  {2}},\ \bibinfo {pages} {024601} (\bibinfo {year} {2018})}\BibitemShut
  {NoStop}%
\bibitem [{\citenamefont {Mravlje}\ \emph {et~al.}(2012)\citenamefont
  {Mravlje}, \citenamefont {Aichhorn},\ and\ \citenamefont
  {Georges}}]{Mravlje12}%
  \BibitemOpen
  \bibfield  {author} {\bibinfo {author} {\bibfnamefont {J.}~\bibnamefont
  {Mravlje}}, \bibinfo {author} {\bibfnamefont {M.}~\bibnamefont {Aichhorn}},\
  and\ \bibinfo {author} {\bibfnamefont {A.}~\bibnamefont {Georges}},\ }\href
  {https://doi.org/10.1103/PhysRevLett.108.197202} {\bibfield  {journal}
  {\bibinfo  {journal} {Phys. Rev. Lett.}\ }\textbf {\bibinfo {volume} {108}},\
  \bibinfo {pages} {197202} (\bibinfo {year} {2012})}\BibitemShut {NoStop}%
\bibitem [{\citenamefont {Pulay}(1980)}]{PULAY1980393}%
  \BibitemOpen
  \bibfield  {author} {\bibinfo {author} {\bibfnamefont {P.}~\bibnamefont
  {Pulay}},\ }\href
  {https://doi.org/https://doi.org/10.1016/0009-2614(80)80396-4} {\bibfield
  {journal} {\bibinfo  {journal} {Chemical Physics Letters}\ }\textbf {\bibinfo
  {volume} {73}},\ \bibinfo {pages} {393 } (\bibinfo {year}
  {1980})}\BibitemShut {NoStop}%
\end{thebibliography}%

\appendix
\section{Finite-size effects}
\begin{figure}[tb]
\includegraphics[width=0.47\textwidth]{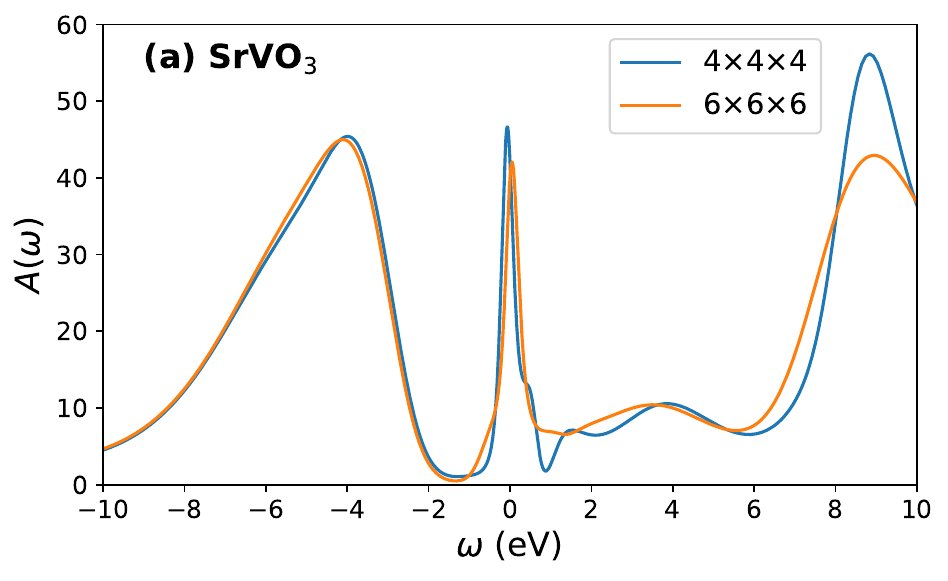}
\includegraphics[width=0.47\textwidth]{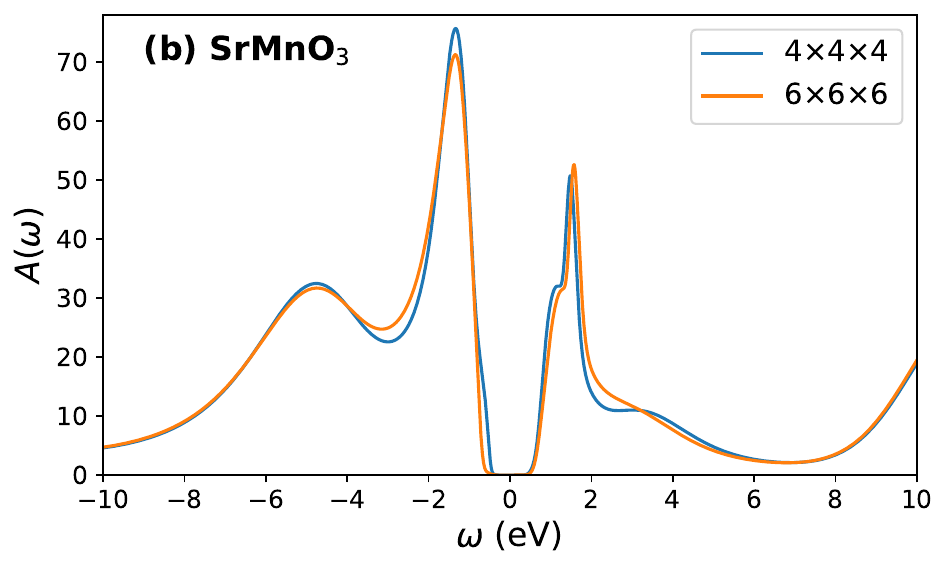}
\caption{Total local DOS for SrVO$_{3}$ and SrMnO$_{3}$ from SEET with different number of k-points in the Brillouin zone. The impurity choices are setup C in Table~\ref{tab:impurities_SrVO3} for SrVO$_{3}$ and setup B in Table~\ref{tab:impurities_SrMnO3} for SrMnO$_{3}$.} \label{fig:finite_size}
\end{figure}

Fig.~\ref{fig:finite_size} shows the finite-size effect for SrVO$_{3}$ and SrMnO$_{3}$. 
From 4$\times$4$\times$4 to 6$\times$6$\times$6, the total local DOS becomes smoother and exhibits fewer features, while the general characteristics remain the same in both Brillouin zone discretizations, for both SrVO$_{3}$ and SrMnO$_{3}$. 

\section{Basis set effect}
\begin{figure}[htp]
\includegraphics[width=0.47\textwidth]{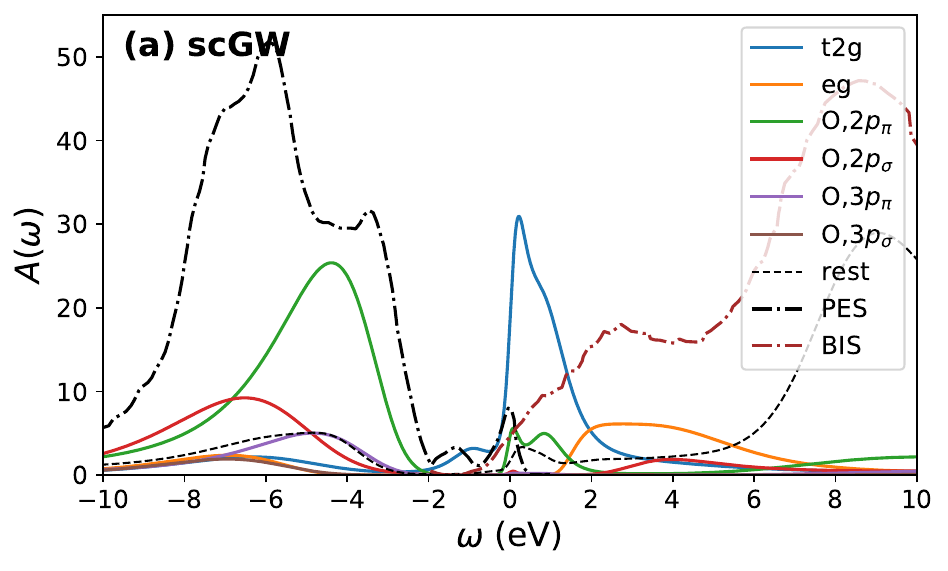}
\includegraphics[width=0.47\textwidth]{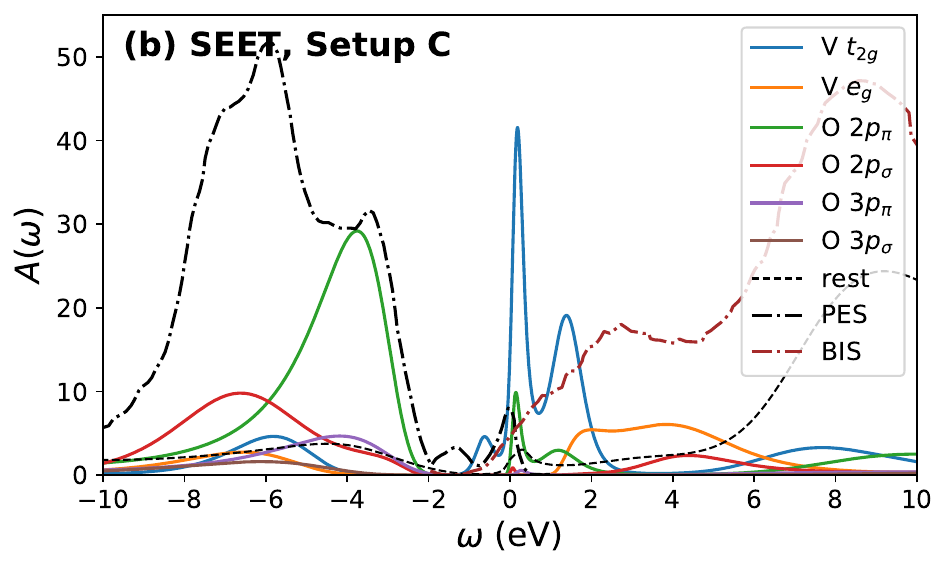}
\caption{Total local DOS for SrVO$_{3}$ from sc$GW$ and SEET. The impurity choices are setup C in Table~\ref{tab:impurities_SrVO3} for SrVO$_{3}$ (The three peak structure for V $t_{2g}$ quasiparticle peak in SEET is due to a coarser k-space discretization ($4\times4\times4$ k-points in the Brillouin zone).} \label{fig:basis_test}
\end{figure}
In order to check whether the missing feature at $\sim$ $-$1.5 eV may be due to a deficiency in our choice of Gaussian basis set, we have repeated the calculation with a larger basis for the vanadium atom in SrVO$_{3}$, \emph{gth-tzvp-molopt-sr}~\cite{GTHBasis}. 
As shown in Fig.~\ref{fig:basis_test}, no qualitative difference was observed for vanadium in the \emph{gth-tzvp-molopt-sr} basis, in both sc$GW$ and SEET calculations.  This indicates that the absence of the feature is unlikely due to a deficiency of the vanadium basis.

\section{$G_{0}W_{0}$ for SrVO$_{3}$}
\begin{figure}[htp]
\includegraphics[width=0.47\textwidth]{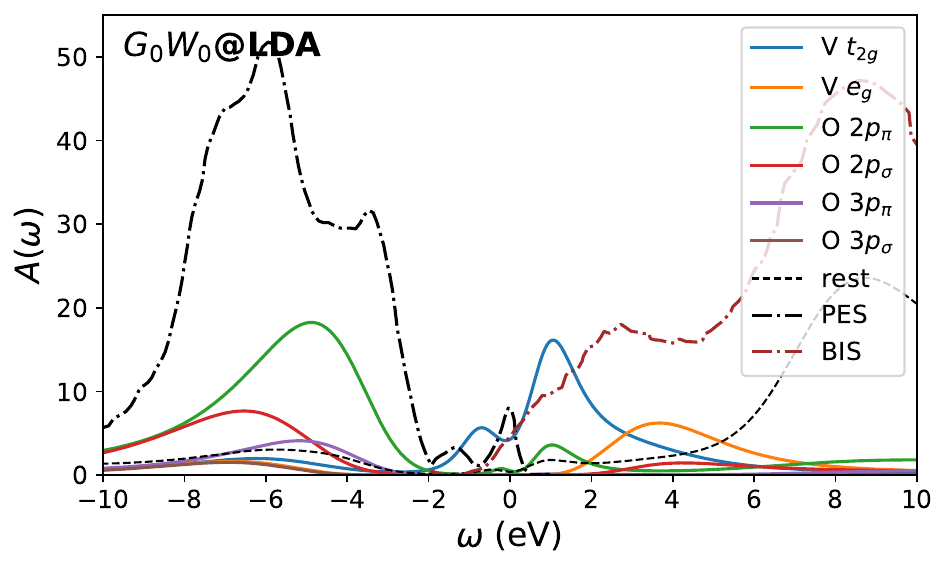}
\caption{Total local DOS for SrVO$_{3}$ evaluated from $G_{0}W_{0}$ based on the LDA band structure.} \label{fig:G0W0_SrVO3}
\end{figure}

Fig.~\ref{fig:G0W0_SrVO3} shows the local DOS evaluated from $G_{0}W_{0}$ based on the LDA band structure.
The $G_{0}W_{0}$ DOS looks qualitatively similar to the sc$GW$ results presented in Fig.~\ref{fig:PDOS_SrVO3} except for a slightly wider $t_{2g}$ quasiparticle bandwidth. 
The hump between $-$1 and $-$2 eV is absent.
Note that the double peak structure is due to the k-space discretization and is not a feature of  a lower satellite.

\section{Convergence of SrMnO$_{3}$}
\begin{figure}[htp]
\includegraphics[width=0.47\textwidth]{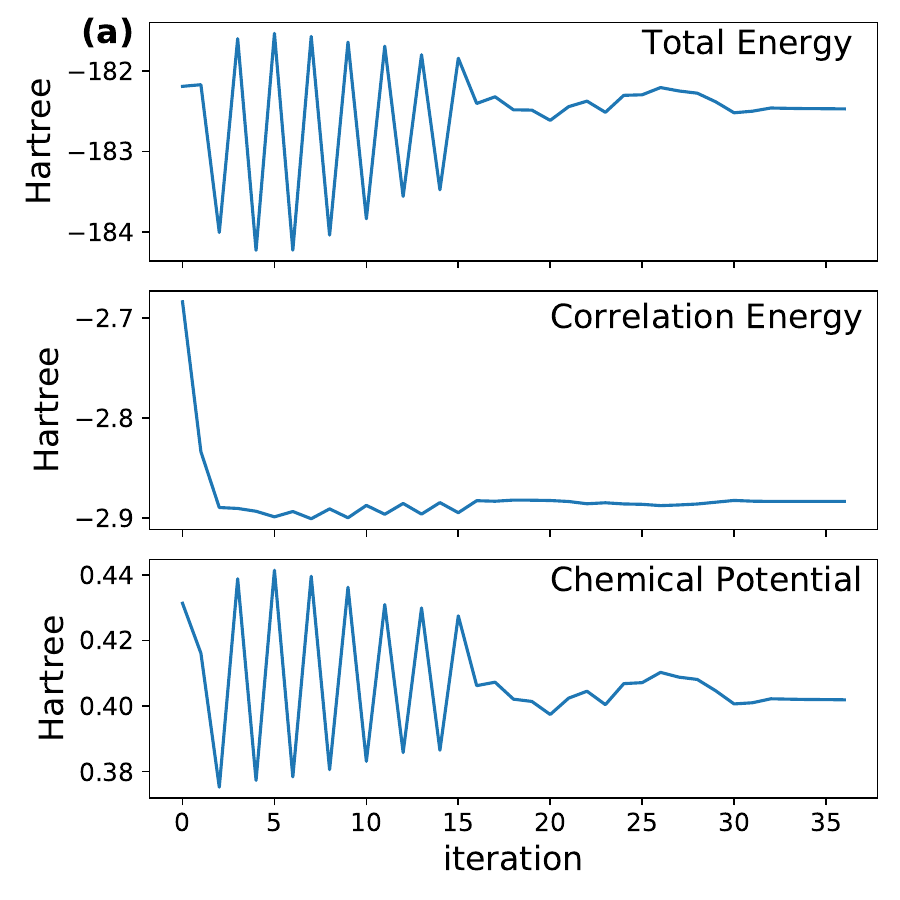}
\includegraphics[width=0.47\textwidth]{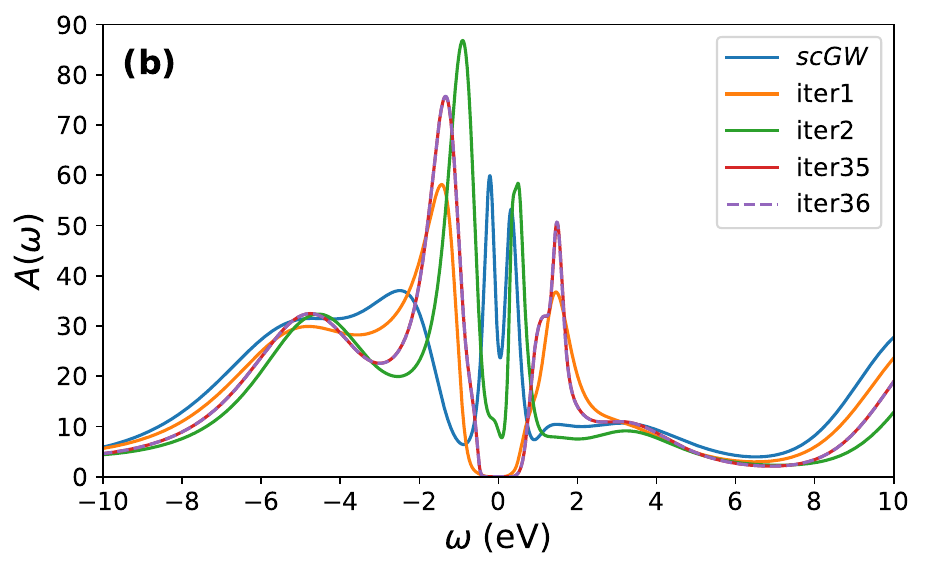}
\caption{Convergence of (a) total energy, correlation energy, chemical potential, and (b) total local DOS for SrMnO$_{3}$ from SEET with impurity choices specified in Setup B.}\label{fig:SEET_convergence_SrMnO3}
\end{figure}

Fig.~\ref{fig:SEET_convergence_SrMnO3} shows SEET convergence of various quantities for SrMnO$_{3}$ with impurity choices specified in Setup B. 
Including local self-energy corrections from Mn $3d$ orbitals opens a gap in the first iteration. The system later oscillates between metallic and insulating states in the following iterations and slowly converged to the physical insulating phase. In order to facilitate the convergence, direct inversion of the iterative subspace (DIIS)~\cite{PULAY1980393} is employed after iteration 17 where much faster convergence is observed. 
Due to it, starting from iteration 17, we observe a much faster convergence.

\section{SEET without outer-loop self-consistency for SrMnO$_{3}$}
\begin{figure}[htp]
\includegraphics[width=0.47\textwidth]{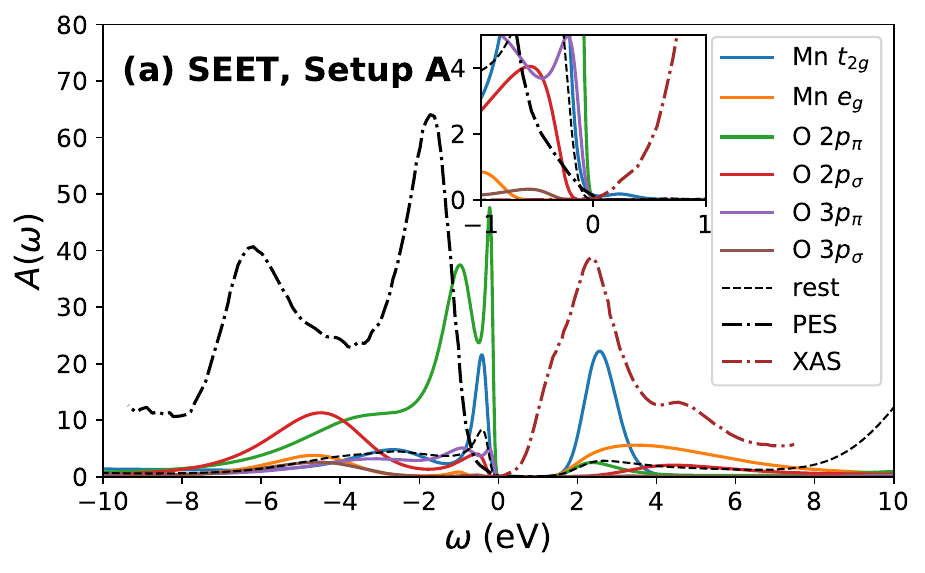}
\includegraphics[width=0.47\textwidth]{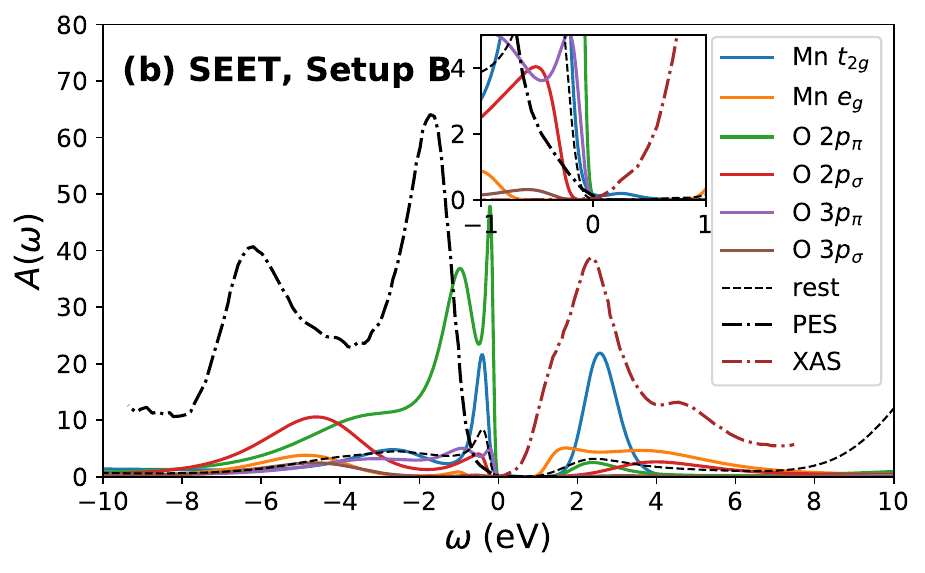} \\
\includegraphics[width=0.47\textwidth]{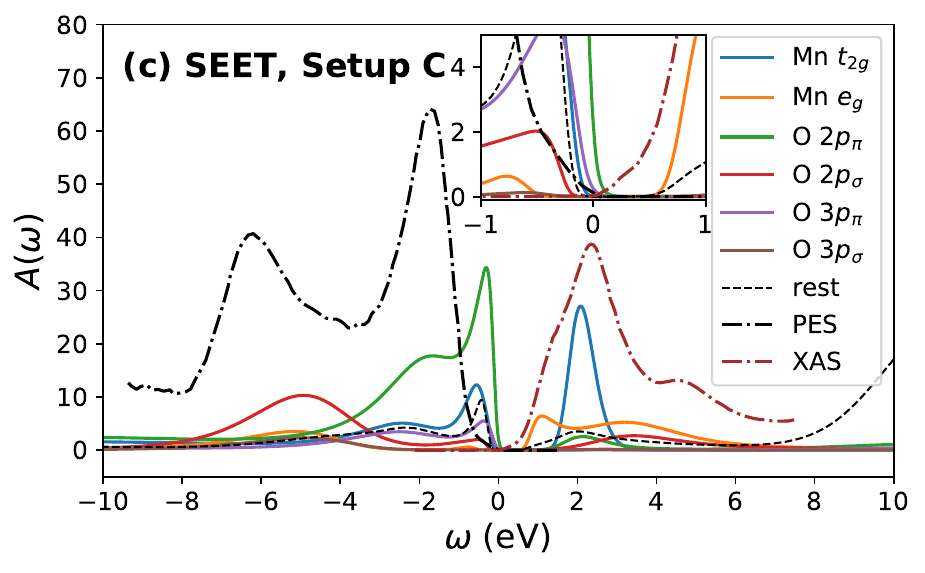}
\caption{Orbital-resolved local DOS for SrMnO$_{3}$ from SEET without an outer-loop self-consistency. The impurity choices from the first to third row correspond to (A), (B), and (C) in Table \ref{tab:impurities_SrMnO3}. The inset shows DOS around $E_{F}$. The dotted lines are photoemission data from ref~ \cite{SrMnFeO3_expt_Kim10}.}\label{fig:SEET_SrMnO3}
\end{figure}

Fig.~\ref{fig:SEET_SrMnO3} shows local orbital-resolved DOS for SrMnO$_{3}$ from SEET without outer-loop self-consistency. 
The impurity choices are listed in Table.~\ref{tab:impurities_SrMnO3}. 
With non-perturbative treatments of the Mn $t_{2g}$ orbitals, the DOS at $E_{F}$ is greatly suppressed while still remaining non-zero (see the inset for Setup A in Fig.~\ref{fig:SEET_SrMnO3}). In addition, an extra sharp quasiparticle peak from O $2p_{\pi}$ orbitals at around $E_{F}$ is observed.
Furthermore including the Mn $e_{g}$ orbitals gives a similar qualitative picture as the one with only Mn $t_{2g}$ orbitals. The only difference arises for the Mn $e_{g}$ conduction band that is pushed forward to $E_{F}$, consistent with results from fully self-consistent SEET.

Next we include O $2p$ orbital in impurity setup C. 
Non-perturbative treatments to O $2p$ states push slightly valence bands away from $E_{F}$, making the sharp quasiparticle peak at $E_{F}$ disappear. 
However, there is still non-zero DOS at $E_{F}$ from O $2p$ states.

In general, although all three impurity choices employed greatly suppress DOS around $E_{F}$ and qualitatively predict correct orbital ordering in both valence and conduction bands, SrMnO$_{3}$ remains metallic with a small  DOS at $E_{F}$. We argue that this unphysical metallic state is an artifact due to the lack of outer-loop self-consistency (the feedback of the strongly correlated orbitals to the weakly correlated part via the Dyson equation). 
Without the outer-loop self-consistency, weakly correlated orbitals outside the impurity subspaces cannot be adjusted by the correlations in the strongly correlated subspaces, resulting in incorrect chemical potential shifts.
\FloatBarrier

\section{SEET total density of states of SrMnO$_{3}$}
\begin{figure}[tbh]
\includegraphics[width=0.47\textwidth]{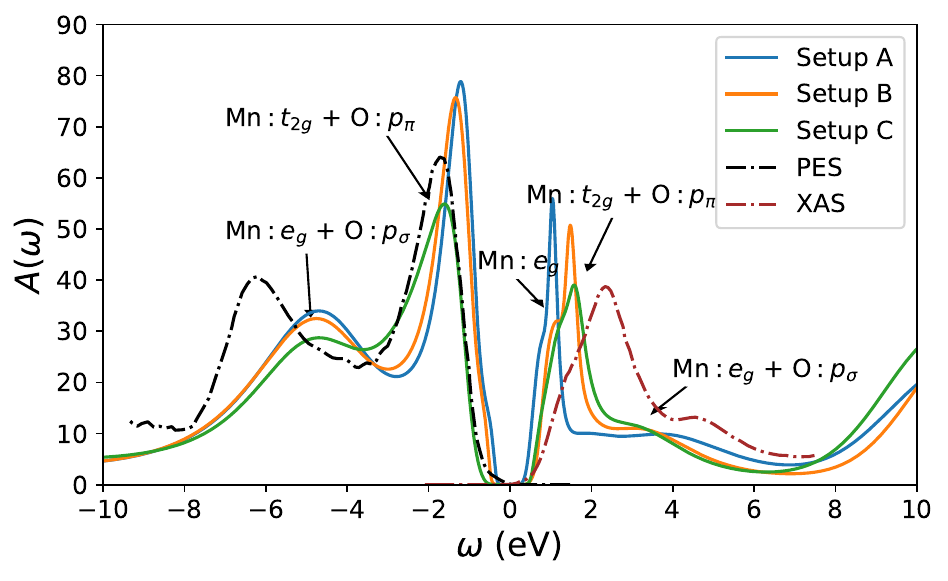}
\caption{Total local density of states of SrMnO$_{3}$ from sc\emph{GW} and SEET with different impurity choices. Arrows indicate the orbital contributions according to SEET with impurity choice C. The PES and BIS data is obtained from Ref.~\cite{SrVO3_PES_Yoshimatsu10,SrVO3_Morikawa95}}\label{fig:SrMnO3_total_DOS}
\end{figure}

Fig.~\ref{fig:SrMnO3_total_DOS} shows the SEET total local DOS of SrMnO$_{3}$. They are obtained by summing all orbitals. We found that the total DOS around the Fermi level is very similar to Fig.~\ref{fig:Mn3d_O2p_vs_Expt}, which shows the sum of the local DOS of the Mn $3d$ + O $2p$ orbitals. 
The quantitative differences come from additional contributions of Mn $4d$ and O $3p$. The rest of orbitals mainly contribute to the lower/higher energy regime far away from the Fermi level. 

\end{document}